\documentclass[11pt]{article}

\usepackage{float}
\usepackage[authoryear,sort&compress,longnamesfirst]{natbib}
\usepackage{setspace}

\usepackage[switch]{lineno}
\usepackage{amssymb, amsmath}
\usepackage{graphicx}
\usepackage{amsthm}
\usepackage{verbatim}
\usepackage{subfig}
\RequirePackage[colorlinks,citecolor=blue,urlcolor=blue]{hyperref}
\usepackage{tikz}
\usetikzlibrary{shapes,arrows}
\usepackage{fancyhdr,booktabs}
\usepackage{setspace}
\usepackage{url}
\usepackage{enumerate}
\newcommand{\del}{\bm{\Delta}}

\usepackage[ruled,lined]{algorithm2e}

\SetKwIF{If}{ElseIf}{Else}{if}{then}{else if}{else}{endif}

\usepackage{amsmath,amsthm,amssymb,fullpage,bm, graphicx, verbatim}
\usepackage{graphicx}

\newtheoremstyle{lemma}
{\topsep} % space above
{\topsep} % space below
{\it} % body fontt
{} % indent
{\bf} % head font in lower caps bolded
{:} % punctuation between head and body
{0.5em} % space after head
{} % manually specify head
\theoremstyle{lemma}

\newtheoremstyle{remark}
{\topsep} % space above
{\topsep} % space below
{} % body font
{} % indent
{\bf} % head font
{:} % punctuation between head and body
{0.5em} % space after head
{} % manually specify head
\theoremstyle{remark}
\newtheorem{rem}{Remark}[section]

\usepackage{tkz-berge}
\usetikzlibrary{fit,shapes}
\usepackage{calc}
\usepackage{tikz}
\usetikzlibrary{decorations.markings}
\tikzstyle{vertex}=[circle, draw, inner sep=0pt, minimum size=6pt]
\newcommand{\ThesisTitle}{A Bayesian Approach to Graphical \\Record Linkage and De-duplication}
\title{\ThesisTitle}
\author{Rebecca C. Steorts$^1$, Rob Hall$^2$, and Stephen E. Fienberg$^3$\\
\\
Department of Statistical Science, Duke University, Durham NC 27707 \\
$^2$Etsy, Inc,
Brooklyn, NY, $^3$Department of Statistics, \\ Carnegie Mellon University, Pittsburgh, PA 15213\\
{beka@stat.duke.edu, rhall@etsy.com, fienberg@andrew.cmu.edu}
\date{\today}
}

\newcommand{\draw}{\stackrel{\text{draw}}{\sim}}

\newcommand{\lam}       {\mbox{$\bm{\Lambda}$}}

\newcommand{\commentt}[1]{}

\newcommand{\by}{\boldsymbol{y}}

\DeclareMathOperator*{\argmax}{arg\,max}

\graphicspath{{images/}}
\DeclareGraphicsExtensions{.pdf}
 
\begin{document}

\pagenumbering{arabic}

\maketitle

\begin{abstract}
  We propose an unsupervised approach for linking records across
  arbitrarily many files, while simultaneously detecting duplicate records
  within files.  Our key innovation involves the representation of  the pattern of links
  between records as a {\em bipartite} graph, in which records are directly
  linked to latent true individuals, and only indirectly linked to other
  records.  This flexible  representation of the linkage structure naturally
  allows us to estimate the attributes of the unique observable people in the
  population, calculate transitive linkage probabilities across records (and represent this visually), and propagate the uncertainty of record linkage into later analyses. Our method makes it particularly easy to integrate record linkage with post-processing procedures such as logistic regression, capture-recapture, etc.  Our linkage structure lends itself to an efficient, linear-time, hybrid Markov chain Monte Carlo algorithm, which overcomes many obstacles encountered by previously record linkage approaches, despite the high-dimensional parameter space. We illustrate our method using longitudinal data from the National Long Term Care Survey and with data from the Italian Survey on Household and Wealth, where we assess the accuracy of our method and show it to be   better in terms of error rates and empirical scalability than other approaches in the literature.
\end{abstract}

\doublespacing
\section{Introduction}
\label{intro}
When data about individuals comes from multiple sources, it is often desirable to match, or link, records from different files that correspond to the same
individual. Other names associated with record linkage are entity
disambiguation, entity resolution, and coreference resolution, meaning that records which are
\emph{linked} or \emph{co-referent} can be thought of as corresponding to the
same underlying \emph{entity}  \citep{christen_2011}.
Solving this 
%long-standing 
problem is not just important as a preliminary step to statistical analysis; the noise and distortions in typical data files make it a difficult, and intrinsically high-dimensional, 
%statistical 
problem \citep{Herzog_2007,lahiri_2005,Winkler_1999,winkler_2000}.

Our methodological advances are in our unified representation of record linkage and de-duplication, via the linkage structure. This lends itself to the use of a large family of models.  The particular one we put forward in this paper is the most basic and minimal member of this family.  We study it not for its realism but for its simplicity and to show what even such a simple model of the family can do. Thus, we propose a Bayesian approach to the record linkage problem based on a
parametric model for categorical data that addresses matching $k$ files simultaneously and includes
duplicate records within lists. We represent the pattern of matches and non-matches as a bipartite graph, in which records are directly linked to the true but latent individuals which they represent, and only indirectly linked to other records.  Such {\em linkage structures} allow us to simultaneously address three problems: record linkage, de-duplication, and estimation of unique observable population attributes.
The Bayesian paradigm naturally handles uncertainty about linkage, which poses a difficult challenge
to frequentist record linkage techniques.
(\citet{liseo_2013} review Bayesian contributions to record linkage).
%Doing so 
A Bayesian approach permits valid inference
 regarding posterior matching probabilities of records and propagation of errors, as we discuss in Section \ref{app}.

To estimate our model, we develop a hybrid MCMC algorithm, in the spirit of \cite{jain_2004}, which runs in linear time
in the number of records and the number of MCMC iterations, even in high-dimensional parameter spaces.
Our algorithm permits duplication across and within lists
but runs faster if there are known to be no duplicates within lists.
We achieve further
gains in speed using standard record linkage blocking techniques \citep{christen_2011}.

We apply our method to data from the National Long Term Care Survey (NLTCS), which tracked and surveyed approximately 20,000 people at five-year intervals.  At each wave of the survey, some individuals had died and were replaced by a new cohort, so the files contain overlapping but not identical sets of individuals, with no within-file duplicates.
We also apply our method to data from the Italian Survey on Household and Wealth (FWIW), a sample survey
383 households conducted by the Bank of Italy every two years. We introduce this application to compare our method to that of \cite{liseo_2011}, whereas applying the latter to the NLTCS study is not computationally feasible in a reasonable amount of time using the competitor's method
as it would take roughly 1500 hours (62 days), whereas our method takes 3 hours.  
%We demonstrate our approach using data from three of the six waves of the National Long Term Care Survey, which tracked and surveyed approximately 20,000 individuals at five-year intervals; these files do not contain duplicates. At each wave of the survey, some individuals had died and were replaced by a new cohort, thus the files contain overlapping but different sets of individuals.
%Furthermore, 
We explore the validity of our method using simulated data. 

Section \ref{motivate} provides a motivating example of the record linkage problem. In Section \ref{model}, we introduce the notation and model, and describe the algorithm in Section \ref{alg}.
Sections \ref{posterior_match} and \ref{sub:function} introduce posterior matching sets which upholding transitivity, and taking functions of the linkage structure. In Sections \ref{app} and \ref{dedup} we apply the method to the the NLTCS under two algorithms (SMERE and SMERED). We also do comparisons, showing that SMERED beats the method of \cite{liseo_2011} for every region of Italy from the FWIW. In Section \ref{user} we review the strengths and limitations of our method, while providing a user's guide for performing record linkage. We evaluate each method compared to a simple baseline and explore the validity of our method under simulation studies. We discuss future directions in Section \ref{disc}.

\subsection{Related Work}
%%TODO: need to unify the Bayesian work and shorten this section.
%%mention larsen and lahiri and larsen and rubin -- propagate error but for only 2 files. most of the Bayesian papers together. 
%%cut in other places
The classical 
work of \cite{fellegi_1969}
considered linking two files in terms of Neyman-Pearson hypothesis testing.  
Compared to this baseline, our approach is distinctive in that it handles multiple files, models distortion explicitly, offers a Bayesian treatment of uncertainty and error propagation, and employs a sophisticated graphical data structure for inference to latent individuals.  
Methods based upon \cite{fellegi_1969} can extend to $k>2$ files \citep{sadinle_multi_1}, but they break down for even 
moderately large $k$
or complex data sets.  Moreover, they 
provide
little information about uncertainty in matches, or about the true values of 
noise-distorted records.
%records which have been corrupted by noise.
\cite{copas_1990} describe the 
idea of modeling the distortion process using 
what they call
%originates
%appears to originate
the ``Hit-Miss Model,'' which
anticipates
 %can be seen as a predecessor of
  part of our model in Section \ref{model}. The specific distortion model we use is, however, closer to that introduced in 
  \cite{hall12}, as part of a nonparametric frequentist technique for matching $k=2$ files that allows for distorted data.  
%  The work of \cite{hall12} introduced a nonparametric frequentist method for matching records in $k=2$ files, allowing for distorted data, 
  Thus, their work is related to ours as we view the records as noisy, distorted entities, that we model using parameters and latent individuals. 
  
There has been much work done in clustering and latent variable modeling in statistics, but also in machine learning and computer science, where of these applications are focused toward author disambiguation. For example, \cite{getoor_2006} proposed a record linkage method  based on latent Dirichlet allocation, which infers the total number of unobserved entities (authors). Their method assumes labeled data such that co-authorship groups can be estimated. In similar work, \cite{dai_2011} used a a non-parametric Dirichlet Process (DP) model, where  groups of authors are associated with topics instead of individual authors. This works well for these specific types of applications such as author disambiguation. However, when clustering records to a hypothesized latent individual, the number of latents typically does not grow as the size of the records does. Hence, a DP process or any non-parametric process tends to over-cluster since we have what we call a \emph{small clustering problem}. Since our goal is to handle a variety of applications including author disambiguation, extensions of our model should be able to handle applications like author disambiguation either using parametric or non-parametric models in sound and principled ways.

Within the Bayesian paradigm,
most work has focused on specialized approaches related to linking two files, which propagate uncertainty \citep{gutman_2013, liseo_2011, larsen_2001, belin_1995, fienberg_1997}. These contributions, while valuable, do not easily generalize to multiple files and to duplicate detection. 
\begin{comment}
\cite{liseo_2011} also 
use
%employ
a parametric Bayesian
method,
%formulation,
but cast record linkage (for $k=2$) as a capture-recapture problem, focusing on
%developing
credible intervals for the unknown population size.  
A key advantage of our approach over theirs is our representation of the linkage structure (Section \ref{notation}): this lets us handle arbitrarily many files, do record linkage, de-duplication and estimation of the unique observable people in the population equally (see Section \ref{app}), and lends itself to our efficient hybrid MCMC scheme.
%A crucial difference between our approach and theirs lies in our representation of the linkage structure (explained in Section \ref{notation}), which allows us to handle arbitrarily many files, and frame the problem as de-duplication, record linkage, estimation of the unknown population size with equal ease.  (We demonstrate how to estimate the unknown population size in Section \ref{app}.)  Our method's hybrid MCMC allows scalability to high-dimensional databases, as opposed to what prior work could handle. 
\end{comment}
%%TODO: make gutman and domingos one paragraph
Three recent papers \citep{domingos_2004, gutman_2013, sadinle_2014} are most relevant to the novelty of our work, namely the linkage structure.
To aid in the recovery of information about the population from distorted records, \cite{gutman_2013} called for developing ``more sophisticated network data structures."  Our linkage graphs are one such  data structure with the added benefit of permitting de-duplication and handling multiple files. Moreover, due to exact error propagation, we  can easily integrate our methods with other
 analytic procedures. Algorithmically, the closest approach to our linkage structure 
is the graphical representation in \cite{domingos_2004}, for de-duplication within one file.  Their representation is a unipartite graph, where records are linked to each other. Our use of a bipartite graph with latent individuals naturally fits in the Bayesian paradigm along with distortion. Our method is the first to handle
%s the only proposed one that handles 
record linkage and de-duplication, while also modeling distortion and running in linear time. 
\textcolor{black}{Finally, \cite{sadinle_2014} recently extended our linkage structure in the representation of a conference matrix, or rather partitioning approaching, where they deviate from our methods using comparison data as to use both categorical and non-categorical data. However, one advantage we maintain is that our methods are more scalable when the data is categorical since a Gibbs sampler does not explore the parameter space as efficiently as a hybrid MCMC approach.}

\section{Motivating Example}
\label{motivate}
The databases (files) contain records regarding individuals that are distorted versions of their unobserved true attributes (fields). We assume that each record corresponds to only one unobserved latent individual. These distortions have various causes---measurement errors, transcription errors, lies, etc.---which we do not model.  We do, however, assume that the probability of distortion is the same for all files (and we do so for computational convenience).  Such distortions, and the lack of unique identifiers shared across files, make it ambiguous when records refer to the same individuals.  This ambiguity can be reduced by increasing the amount of information available, either by adding informative fields to each record, or, sometimes, by increasing the number of files.

We illustrate this issue with a motivating example of real world distortion and noise (see Table~\ref{real}). With gender and state alone, these records could refer to (i) a single individual with a data entry error (or old address, etc.) in one file; (ii) one individual correctly recorded as living in SC and another correctly recorded in WV; (iii) two individuals with errors in at least one address; (iv) three distinct individuals with correct addresses; (v) three individuals with errors in addresses.  (There are still further possibilities if the gender field might contain errors.)  The goal is to determine whether distinct records refer to the same individual, or to distinct individuals.

Table~\ref{real} illustrates a scenario where there is considerable
uncertainty about whether two records correspond to the same individual (under just gender and state).  
As we mentioned earlier, the identities of the true individuals to which the records correspond are not entirely clear due to the limited information available.
Suppose we expand the field information 
by adding date of birth (DOB) and race.
We still have the same host of possibilities as before, but the addition of DOB \emph{may} let us make better decisions about matches.
It is not clear if File 1 and File 2 are the same person or different people (who just happen to have the same birthdate). However, the introduction of DOB does make it more likely that File 3 is not the same person as in File 1 and File 2. The method we propose in Section~\ref{model} deals with this type of noise; however, it proposes to deal with noisier records in that they traditionally do not have identifying information such as name and address, making the matching problem inherently difficult, even indeterminate.

\begin{table}[htbp]
\begin{center}
\begin{tabular}{l|lc|ccc}
%&Name&Address&Year\\ \hline
&Gender & State&DOB & Race\\ \hline
%File~1& \textcolor{red}{J}. Smith& 1320 \textcolor{red}{East} Liberty \textcolor{red}{Dr.} & \textcolor{blue}{1983}\\
%File~2& \textcolor{red}{Jon} Smith& 1320 Liberty \textcolor{red}{St.} & \textcolor{blue}{1983}\\
%File~3& \textcolor{red}{John} Smith& 1320 Liberty \textcolor{red}{St.} & \textcolor{blue}{1943}
File~1 &  F  & SC & \textcolor{blue}{04/15/83} & \textcolor{blue}{White} \\
File~2 & F  &  \textcolor{red}{WV}  & \textcolor{blue}{04/15/83} & \textcolor{blue}{White} \\
File~3 & F & SC & \textcolor{blue}{07/25/43}  & \textcolor{blue}{White} 
\end{tabular}
\caption{Three files  with year of birth and race.}
\label{real}
\end{center}
%\label{default}
\end{table}%

\section{Notation, Assumptions, and Linkage Structure}
\label{notation}
We begin by defining some notation,
for $k$ files or lists. 
\textcolor{black}{For simplicity, we assume that all files
contain $p$ fields in common, which are all categorical, field $\ell$ having
$M_{\ell}$ levels. Thus, we do not handle missingness of fields across databases, however when we do have $p$ fields in common, some information could be missing.}
% \textcolor{red}{We also assume that every database has a minimal number of overlapping fields for each record, such that all the databases can be merged. In essence,
%we assume there exists $f$ overlapping fields (and hence, we do not consider the case where entire fields in databases are missing).
%Within the fields that do overlap, there still may be missing values. }
Handling
missing-at-random fields within records is a minor extension within the
Bayesian framework \citep{reiter_2007}.
Let $\bm{x}_{ij}$ be the data for the $j$th record in file $i$, where
$i=1,\ldots,k$,\; $j=1,\ldots,n_i$, and $n_i$ is the number of records in file
$i$; $\bm{x}_{ij}$ is a categorical vector of length $p$.  Let $\bm{y}_{j'}$
be the latent vector of
%unobserved
true field
values
% records
for the $j'$th individual in the population (or rather aggregate sample), where
$j'=1,\ldots,N$, 
$N$ being
% and $N$ denotes
the total number of \emph{observed}
individuals from the population. 
%Note that
$N$ could be as small as 1 if every
record 
in every file
refers to the same individual
% in all $k$ datasets
or as large as
$N_{\max} \equiv \sum_{i=1}^k{n_i}$ if no datasets share any
individuals.
%Since $N$ is unknown, for convenience, we take it to be the upper bound.

Next, we define the linkage structure
$\bm{\Lambda}=\{\lambda_{ij}\;;\;i=1,\ldots,k\;;\;j=1,\ldots,n_i\}$ where
$\lambda_{ij}$ is an integer from $1$ to $N_{\max}$ indicating which latent
individual the $j$th record in file $i$ refers to, i.e., $\bm{x}_{ij}$ is a
possibly-distorted measurement of $\bm{y}_{\lambda_{ij}}$.
Finally,
$z_{ij\ell}$ is $1$ or $0$ according to whether or not a particular
field $\ell$ is distorted in $\bm{x}_{ij}.$

As usual, we use $I$ for indicator functions (e.g., $I(x_{ij\ell}=m)$ is 1 when
the $\ell$th field in record $j$ in file $i$ has the value $m$), and $\delta_a$ for the distribution of a point mass at $a$ (e.g., $\delta_{y_{\lambda_{ij}\ell}}$).
The vector~$\bm{\theta}_{\ell}$ of length $M_{\ell}$ denotes the multinomial
probabilities. 
%
%For example, if we think each sex is equally likely, then $\bm{\theta}_{\ell} = (0.5,0.5).$
%For example, if $\theta_\ell=(0.5,0.5)$, then each sex is equally likely.
For clarity, we always index as follows:
%For clarity, we index as follows below:
${i=1,\ldots,k;}$ ${j=1,\ldots,n_i;}$ ${j'=1,\ldots,N;}$ ${\ell=1,\ldots,p;}$ ${m=1, \ldots, M_{\ell}.}$
We provide an example of the linkage structure and a graphical representation in Appendix \ref{example}. 

%\subsection{A Bayesian Parametric Model of Independent Fields}
\subsection{Independent Fields Model}
\label{model}

 %
%In the notation introduced in Section~\ref{notation},
We assume the $k$ files are conditionally independent, given the latent
individuals, and that fields are independent within individuals (this is done for computational simplicity as is the motivation for our Bayesian model). 
We formulate the following Bayesian parametric model:
% where we sample from the full conditionals
%using a hybrid MCMC algorithm:
%
\begin{align*}
\bm{x}_{ij\ell}\mid\lambda_{ij},\bm{y}_{\lambda_{ij}\ell},z_{ij\ell},\bm{\theta}_\ell&\stackrel{\text{ind}}{\sim}
\begin{cases}
\delta_{\bm{y}_{\lambda_{ij}\ell}}&\text{ if }z_{ij\ell}=0\\
\text{MN}(1,\bm{\theta}_\ell)&\text{ if }z_{ij\ell}=1
\end{cases}\\
z_{ij\ell}&\stackrel{\text{ind}}{\sim}\text{Bernoulli}(\beta_\ell)\\
\bm{y}_{j'\ell}\mid\bm{\theta}_{\ell}&\stackrel{\text{ind}}{\sim}\text{MN}(1,\bm{\theta}_\ell)\\
\bm{\theta}_\ell&\stackrel{\text{ind}}{\sim}\text{Dirichlet}(\bm{\mu}_\ell)\\
\beta_\ell&\stackrel{\text{ind}}{\sim}\text{Beta}(a_\ell,b_\ell) \\
\pi(\lam) &\propto 1,
\end{align*}
where
$a_\ell,b_\ell,$ and $\bm{\mu}_\ell$ are all known.

\begin{rem}
  We assume that every legitimate configuration of the $\lambda_{ij}$ is equally
  likely a~priori.
% i.e., $\pi(\lam) \propto 1$
  This implies a
  % corresponds to a
  non-uniform prior on
  % other
  related quantities,
  % of interest, however,
  such as the number of individuals in the data.  
  %Nonetheless, 
  The uniform
  prior on $\bm{\Lambda}$ is convenient, 
  since it simplifies computation of the posterior.  Devising non-uniform priors over linkage structures remains a challenging problem both computationally and statistically, as sensible priors must remain invariant when permuting the labels of latent individuals, and cannot use covariate information about records.
  
%  it is not immediately clear how to  construct either a subjective
%  or an alternative objective prior.
  % We currently take the uniform prior on $\bm{\Lambda}$ purely out of
  % convenience, since construction of either a subjective or an alternative
  % objective prior is extremely complex.
\end{rem}
Deriving the joint posterior and conditional distributions is mostly
straightforward.  One subtlety, however, is that $\bm{y}$, $\bm{z}$ and
$\bm{\Lambda}$ are all related, since if $z_{ij\ell}=0$, then it must be the
case that $y_{\lambda_{ij}\ell}=x_{ij\ell}$.
Taking this into account,
% we find that
the joint posterior is
%\begin{align*}
%\label{joint}
%&\pi(\bm{\Lambda},\bm{y},\bm{z},\bm{\theta},\bm{\beta}\mid\bm{x})\\
%&\propto
%\prod_{i,j,\ell, m}
%%\prod_{i=1}^{k}\prod_{j=1}^{n_i}\prod_{\ell=1}^p\prod_{m=1}^{M_\ell}
%\left[(1-z_{ij\ell})\delta_{y_{\lambda_{ij}\ell}}(x_{ij\ell})
%\;\;+\;\;
%z_{ij\ell}\theta_{\ell m}^{I(x_{ij\ell} = m)}\right]\notag\\
%&\phantom{\propto{}}\times
%\prod_{\ell, m}
%%\prod_{\ell=1}^p\prod_{m=1}^{M_\ell}
%\theta_{\ell m}^{\mu_{\ell m}+\sum_{j'=1}^N I(y_{j'l}=m)}\notag\\
%&\phantom{\propto{}}\times
%\prod_{\ell}
%%\prod_{\ell=1}^p
%\beta_\ell^{a_\ell-1+\sum_{i=1}^k\sum_{j=1}^{n_i}z_{ij\ell}} \notag \\
%&\times
%(1-\beta_\ell)^{b_\ell-1+\sum_{i=1}^k\sum_{j=1}^{n_i}(1-z_{ij\ell})}. \notag
%\end{align*}
%
\begin{align*}
%\label{joint}
\pi(\bm{\Lambda},\bm{y},\bm{z},\bm{\theta},\bm{\beta}\mid\bm{x})
&\propto
\prod_{i,j,\ell, m}
%\prod_{i=1}^{k}\prod_{j=1}^{n_i}\prod_{\ell=1}^p\prod_{m=1}^{M_\ell}
\left[(1-z_{ij\ell})\delta_{y_{\lambda_{ij}\ell}}(x_{ij\ell})
\;\;+\;\;
z_{ij\ell}\theta_{\ell m}^{I(x_{ij\ell} = m)}\right]\notag\\
& \qquad \times
\prod_{\ell, m}
%\prod_{\ell=1}^p\prod_{m=1}^{M_\ell}
\theta_{\ell m}^{\mu_{\ell m}+\sum_{j'=1}^N I(y_{j'l}=m)}
\times
\prod_{\ell}
%\prod_{\ell=1}^p
\beta_\ell^{a_\ell-1+\sum_{i=1}^k\sum_{j=1}^{n_i}z_{ij\ell}} \notag \\
& \qquad \times
(1-\beta_\ell)^{b_\ell-1+\sum_{i=1}^k\sum_{j=1}^{n_i}(1-z_{ij\ell})}. \notag
\end{align*}
%
%We suppress
%derivation of the full conditionals, but note that the full conditionals of 
%$\bm{y}$,
%$\bm{z}$ and $\bm{\Lambda}$ always obey their logical dependence, and therefore never condition on impossible events.
%The full conditional of $\bm{\Lambda}$ must reflect whether or not
%there are duplicates within files.  If we define $R_{ij^{\prime}} = \left\{ j : \lambda_{ij} = j^{\prime}\right\},$ then not having within-file duplicates means that $R_{ij^\prime}$ must be either $\emptyset$ or a single record, for each $i$ and $j^{\prime}$. Graphically, this means
%%can be seen as 
%allowing
%or forbidding links from a latent individual to multiple records within one
%file. 
Now we consider the conditional distribution of $\bm{y}.$ Here, the part of the
posterior involving~$\bm{x}$ only matters for the conditional of~$\bm{y}$ when
$z_{ij\ell}=0$.  Specifically, when $z_{ij\ell}=0$, we know that
$y_{\lambda_{ij}\ell}=x_{ij\ell}.$ 
Next, for each $j'=1,\ldots,N$, let $R_{ij'}=\{j:\lambda_{ij}=j'\}$, so that $\bm{x}_{ij}$ and $\bm{y}_{j'}$ refer to the same individual if and only if $j\in R_{ij'}$.  

\begin{rem}
This notation allows the consideration of duplication within
lists, i.e., distinct records within a list that correspond to the same
individual.  In particular, two records $j_1$ and~$j_2$ in the same list~$i$
correspond to the same individual if and only if
$\lambda_{ij_1}=\lambda_{ij_2}.$  Implementing this in our hybrid MCMC is {\em simpler} than assuming the lists are already de-duplicated, since de-duplication implies that certain linkages are undefined.
\end{rem}

From the joint posterior above, we can write down the full conditional
distributions of $\bm{\theta}$ and $\bm{\beta}$ directly as
\begin{align*}
%\label{beta}
\beta_\ell \mid \bm{\Lambda},\bm{z},\bm{\theta}, \by ,\bm{x}
\sim \text{Beta}\left(a_{\ell} + \sum_{i=1}^k \sum_{j=1}^{n_i} z_{ij\ell} , b_{\ell} + \sum_{i=1}^k \sum_{j=1}^{n_i} (1-z_{ij\ell}) \right)
\end{align*}
 for all $\ell$ and  
\begin{align*}
%\label{theta}
\theta_{\ell m} \mid   \bm{\Lambda}, \bm{z}, \by, \bm{\beta},\bm{x} 
\sim 
\text{Dirichlet}\left(\mu_{\ell m} + \sum_{j'=1}^N y_{j'\ell }  + \sum_{i=1}^k \sum_{j=1}^{n_i}   z_{ij\ell}\; x_{ij\ell} + 1\right),
\end{align*}
 for all $\ell$ and for each $m=1,\ldots, M_{\ell}.$
Then
\begin{align*}
\bm{y}_{j'l}\mid\bm{\Lambda},\bm{z},\bm{\theta},\bm{\beta},\bm{x}
&\sim\begin{cases}
\delta_{x_{ij\ell}}
%\text{ for every }j\in R_{ij'}
&\text{if there exist } i,j\in R_{ij'}\text{ such that }z_{ij\ell}=0,\\
\text{Multinomial}(1,\bm{\theta}_{l})&\text{otherwise}.
\end{cases}
\end{align*}
In other words, the linkage structure $\bm{\Lambda}$ tells us that
$\bm{y}_{j'}$ corresponds to some $\bm{x}_{ij}$ when there is no distortion. 
Now consider that $\bm{z}$ represents the indicator of whether or not there is
a distortion.  When we condition on $\bm{x},\bm{y}$, and $\bm{\Lambda}$, there
are times when a distortion is certain. Specifically, if $x_{ij\ell}\ne y_{\lambda_{ij}\ell},$ then we know there must be a distortion, so $z_{ij\ell}=1.$
%Note that $\delta_1$ is just the same thing as Bernoulli(1).
However, if $x_{ij\ell}=y_{\lambda_{ij}\ell}$, then $z_{ij\ell}$ may or may not equal 0. 
Therefore, we
can show
\begin{align*}
&P(z_{ij\ell}=1\mid\bm{\Lambda},\bm{y},\bm{\theta},\bm{\beta},\bm{x})\\
&=\frac{P(\bm{\Lambda},\bm{y},\bm{\theta},\bm{\beta},\bm{x}\mid z_{ij\ell}=1)P(z_{ij\ell}=1)}
{P(\bm{\Lambda},\bm{y},\bm{\theta},\bm{\beta},\bm{x}\!\mid\!z_{ij\ell}=1)P(z_{ij\ell}=1)+P(\bm{\Lambda},\bm{y},\bm{\theta},\bm{\beta},\bm{x}\!\mid\!z_{ij\ell}=0)P(z_{ij\ell}=0)}\\
&=\frac{\beta_\ell\prod_{m=1}^{M_\ell}\theta_{\ell m}^{x_{ij\ell}}}
{\beta_\ell\prod_{m=1}^{M_\ell}\theta_{\ell m}^{x_{ij\ell}}+(1-\beta_\ell)}.
\end{align*}
%Why is there nothing multiplied by the $1-\beta_\ell$ above?  Well, if $z_{ij\ell}=0$, then there is no distortion, and we're guaranteed to get $x_{ij\ellm}$ equal to $y_{\lambda_{ij}\ellm}$ for each~$m$.
Then we can write the conditional as
\begin{align*}
&z_{ij\ell}\mid\bm{\Lambda},\bm{y},\bm{\theta},\bm{\beta},\bm{x}\stackrel{\text{ind}}{\sim}\text{Bernoulli}(p_{ij\ell}),\text{ where }\\
&p_{ij\ell}=\begin{cases}
1&\text{ if }x_{ij\ell}\ne y_{\lambda_{ij}\ell}\\
\dfrac{\beta_\ell\prod_{m=1}^{M_\ell}\theta_{\ell m}^{x_{ij\ell}}}
{\beta_\ell\prod_{m=1}^{M_\ell}\theta_{\ell m}^{x_{ij\ell}}+(1-\beta_\ell)}
&\text{ if }x_{ij\ell}=y_{\lambda_{ij}\ell},
\end{cases}
\quad \text{for all } \ell.
\end{align*}

Finally, we derive the conditional distribution of $\bm{\Lambda}$.  This is the only part of the model which changes if we allow duplication within lists. Conditional on
$\bm{x},\bm{y}$, and $\bm{z}$, we can rule out
there are many linkage structures , i.e., ones that have probability zero.  
Specifically, for
any $i,j,\ell$ such that $z_{ij\ell}=0$, there is no distortion.  This
means that if $z_{ij\ell}=0$, then for any~$c$ such that
$x_{ij\ell}\ne y_{cl}$, we know that $\lambda_{ij}=c$ is impossible.
On the other hand, if $z_{ij\ell}=1$, then $\bm{x}_{ij\ell}$ simply comes from a
multinomial, in which case the linkage structure is totally
irrelevant. %There is another restriction on $\lambda_{ij}$ as well.
If we assume that no duplication is allowed within each list, then $j_1\ne j_2$ implies that
$\lambda_{ij_1}\ne\lambda_{ij_2}$. 
%\textcolor{red}{%{[we've gotten rid
%of this restriction if we consider duplicates]}\\
%%
%
Additionally, we note that for each file~$i,$ the part of the
  linkage structure corresponding to that dataset
  ($\lambda_{i1},\ldots,\lambda_{in_i}$) is independent of the linkage
  structures for the other datasets conditional on everything else.
 Then we can write the following (assuming no duplicates):
\begin{align*}
P(\lambda_{i1}=c_1,\ldots,\lambda_{in_i}=c_{n_i}\mid\bm{y},\bm{z},\bm{\theta},\bm{\beta},\bm{x})
&\stackrel{\text{ind}}{\propto}
\begin{cases}
0&\text{ if there exist }j,\ell\text{ such that }\\
&\qquad z_{ij\ell}=0 \text{ and }x_{ij\ell}\ne y_{c_j \ell},\\
&\qquad\text{or if } c_{j_1}=c_{j_2} \text{ for any } j_1\ne j_2\\
1&\text{ otherwise, }
\end{cases}
\end{align*}
where the somewhat nonstandard notation $\stackrel{\text{ind}}{\propto}$ simply
denotes that distributions for different~$i$ are independent.

Allowing duplicates within lists, we find that the conditional distribution lifts a restriction from the one just derived.
That is, 
\begin{align*}
P(\lambda_{i1}=c_1,\ldots,\lambda_{in_i}=c_{n_i}\mid\bm{y},\bm{z},\bm{\theta},\bm{\beta},\bm{x})
&\stackrel{\text{ind}}{\propto}
\begin{cases}
0&\text{ if there exist }j,\ell\text{ such that }\\
&\qquad z_{ij\ell}=0 \text{ and }x_{ij\ell}\ne y_{c_j \ell},\\
1&\text{ otherwise.}
\end{cases}
\end{align*}

\subsection{Split and MErge REcord linkage and De-duplication (SMERED)
Algorithm}
\label{alg}

Our main goal is estimating the posterior distribution of the linkage (i.e.,
the clustering of records into latent individuals). The simplest way of accomplishing
this is via Gibbs sampling.  We could iterate through the records, and for each
record, sample a new assignment to an individual (from among the individuals
represented in the remaining records, plus an individual comprising only that
record).  However, this requires the quadratic-time checking of proposed
linkages for every record.
%requires checking the legality of proposed linkages for
%every record, which is a quadratic-time task,
%is at least quadratic in time,
%\footnote{For each record $i,j$, propose replacing $\lambda_{ij}$ with a random value in $1$ to $N_{\max}$; if this is legal, given all other values in $\bm{\Lambda}$ and $\bm{X}$, $\bm{y}$, $\bm{z}$, accept, otherwise reject.  Checking legality takes time $O(n_i)$ for record $i,j$, so the over-all time would be $O(N_{\max} n_i)$ which is at worst $O(N_{\max}^2).$}
%
%as we show below, (Section \ref{alg}) at worst linear-time.
Thus, instead of  Gibbs sampling, we use a hybrid MCMC algorithm to explore the space of possible linkage structures, which allows our algorithm to run in linear time.
%Modifications needed to allow duplication within lists will be explained as they arise.

%In
Our hybrid MCMC
takes
% we take
advantage of split-merge moves, as done in \cite{jain_2004},
% as well as standard record blocking heuristics,
which avoids the problems associated with
Gibbs sampling,
even though the number of parameters grows with the number of records.
% since our parameter space is high-dimensional as the number of records increase.
This is accomplished via proposals that can traverse the state space quickly
and frequently visit high-probability modes, since the algorithm splits or
merges records in each update, and hence, frequent updates of the Gibbs sampler
are not necessary.

\textcolor{black}{Blocking is a common technique that creates places similar records into partitions or blocks based on some rule or automated process. For a review of cladssical and newer blocking methods see \citep{winkler_2000, steorts_2014_hash}.
The form of blocking that we illustrate in our examples requires an exact match
in certain fields (e.g., birth year) if records are to be
% identified as
linked.
\emph{Blocking}
%, called blocking,
can greatly reduce the number of possible 
links
% linkages
between records.
In our application, we use an approximate blocking procedure, not an exact one.
%The methods in our paper use an approximate blocking procedure, not exact ones, as we illustrate in the applications section. 
%Our algorithm can also handle any type of blocking, where the blocking are independent, such as those considered in \cite{steorts_2014_hash}, however, we 
%look at simple forms of blocking here since more complicated forms are not needed in both applications. 
}
Since blocking gives up on finding truly co-referent records which disagree
on those fields, it is best to block on
fields that have little or no distortion.
We block on the fairly reliable fields of sex and
birth year in our application to the NLTCS below.
%There is a tradeoff made in assume blocking, in that we give
%up the possibility of making links between records in gains in speed, where we
% block on records that are thought to have no or very low distortion. 
A strength of
our model
% the Bayesian model presented here
is that it 
incorporates blocking organically.
%can incorporate the idea of blocking quite organically. 
Setting
%If we take
$b_\ell=\infty$ for a particular field $\ell$
forces
% then we force
the distortion probability for 
that field to zero.
%field~$\ell$ to be zero.
This requires 
matching
%co-referent
records to agree on the $\ell$th field, 
just like blocking.
% thereby enforcing a blocking restriction.

We now discuss how the split-merge process links records to records, which it
does by
assigning
% assignment of
records to latent individuals.  Instead of sampling 
assignments
%the assignments of records
at the record level, we do so at the individual level.  Initially, each record
is assigned to a unique individual.  On each iteration,
% of MCMC,
we
% repeatedly
choose two records at random.  If 
the pair
%those two records
belong to 
\emph{distinct}
% two \emph{different}
latent individuals, then we propose merging those individuals to form a single
new latent individual (i.e., we propose that those records are co-referent).
On the other hand, if the two records belong to the \emph{same} latent
individual, then we propose splitting it into two new latent individuals, each
seeded with one of the two chosen records, and the other records randomly
divided between the two.  Proposed splits and merges are accepted based on the
Metropolis-Hastings ratio and rejected otherwise.

\textcolor{black}{Sampling
% uniformly
from all possible pairs of records will sometimes lead to proposals to merge
records in the same list.  If we permit duplication within lists, then this is
not a problem.  However, if we know (or assume) there are no duplicates within
lists, we should avoid wasting time on such pairs.
%enforce an assumption of no duplication within
%lists, then we should avoid sampling pairs of records within the same list.
The no-duplication version of our algorithm does precisely this. \textcolor{black}{(See Appendix \ref{algorithm} for the algorithm and pseudocode.)} \textcolor{black}{When there are no duplicates within files, we call this the SMERE (Split and MErge REcord linkage) algorithm, which enforces the restriction that $R_{ij^{\prime}}$ must be either $\varnothing$ or a single record.  This is done through limiting the proposal of record pairs to those in distinct files; the algorithm otherwise matches SMERED.} }
%Note: Our algorithm is linear in the number of records and MCMC iterations, which we show in Appendix \ref{time}. 

\subsection{Posterior Matching Sets and Linkage Probabilities}
\label{posterior_match}
In a Bayesian framework, the output of record linkage is not a deterministic set of matches between records, but a probabilistic description of how likely records are to be co-referent, based on the observed data.
Since we are linking multiple files at once, we propose a range of
posterior matching probabilities:
%In the absence of unique identifiers, we would be interested in an unsupervised
%approach, which is why we propose posterior matching sets and linkage
%probabilities of records. Hence, we discuss a range of posterior matching
%probabilities:
the posterior probability of linkage
between
two arbitrary records and more generally
among
$k$ records, the posterior probability that a given set of records is
linked, and the posterior probability that a given set of records is a maximal
matching set (which will be defined later). \textcolor{black}{Furthermore, we make a connection to \cite{liseo_2011} of why our proposed posterior matching probabilities our \emph{optimal}.}

%We say that
Two records $(i_1,j_1)$ and $(i_2,j_2)$ {\em match} if they
point to the same latent individual, i.e., if $\lambda_{i_1j_1} = \lambda_{i_2j_2}.$
%In our Bayesian framework, matches are probabilistic rather than deterministic,
%and
The posterior probability of a match can be computed from the $S_G$ MCMC
samples:
%A posterior probability of a match between two records $j_1$ and $j_2$ is simply the Bayes estimate of the indicator of the match under squared error loss. 
%For example, the posterior probability of a match between any pair of records 
%$((i_1,j_1),(i_2,j_2)),$ is 
$$
P(\lambda_{i_1j_1} = \lambda_{i_2j_2} | \bm{X}) = \frac{1}{S_G}\sum_{h=1}^{S_G}
I(\lambda_{i_1j_1}^{(h)} = \lambda_{i_1j_2}^{(h)}).$$ 
A one-way match
occurs
when an individual appears in only one of the $k$ files, while
% whereas
a two-way match
is
%refers to
when an individual appears in exactly two of the $k$ files, and so on (up to $k$-way
matches).  We approximate the posterior probability of arbitrary one-way,
two-way, \ldots, $k$-way matches as the
ratio of
frequency of those matches in the posterior sample
to $S_G$.
%, over $H$.

\textcolor{black}{Although probabilistic results and interpretations provided by the Bayesian paradigm are useful both quantitatively and conceptually, 
we often need to report a 
point estimate of the linkage structure.  
%``answer,'' i.e., a single overall linkage structure that we feel is most probable based on the data.  This scenario is analogous to that of any simple quantitative  statistical problem in which a single-number point estimate is desired as a ``best guess'' of an unknown parameter. 
 Thus, we face the question of how to condense the overall posterior distribution of $\bm\Lambda$ into a single estimated linkage structure.}

\textcolor{black}{Perhaps the most obvious approach is to set some threshold~$v$, where $0<v<1$, and to declare (i.e., estimate) that two records match if and only if their posterior matching probability exceeds~$v$.  This strategy is useful if only a few specific pairs of records are of interest, but its flaws are exposed when we consider the coherence of the overall estimated linkage structure implied by such a thresholding strategy.  Note that the true linkage structure is \emph{transitive} in the following sense: if records A and B are the same individual, and records B and C are the same individual, then records A and C must be the same individual as well.  This requirement of transitivity, however, is in no way enforced by the simple thresholding strategy described above.  Thus, a more sophisticated approach is required if the goal is to produce an estimated linkage structure that preserves transitivity.}

\textcolor{black}{To this end, it is useful to define a new concept.}
%We define a
A set of records $\mathcal{A}$
is
% to be
a \emph{maximal matching} set (MMS) if
\textcolor{black}{every record in the set has}
the same value of ${\lambda}_{ij}$ and
\textcolor{black}{no record outside the set has that}
value of ${\lambda}_{ij}.$ Define $\bm{\Omega}(\mathcal{A}, \bm{\Lambda})$ to be
%$I(\mathcal{A}\text{ is a maximal matching set in }\bm{\Lambda}), $ which is
1 if 
$\mathcal{A}$ is
%we have
an MMS in $\bm{\Lambda}$ and 0 otherwise:
$$\bm{\Omega}(\mathcal{A}, \bm{\Lambda}) = \sum_{j^{\prime}}{\left( \prod_{(i,j) \in \mathcal{A}}{I(\lambda_{i j}=j^{\prime})}\prod_{(i,j)\not\in\mathcal{A}}{I(\lambda_{ij}\neq j^{\prime}})\right)}.$$

Essentially, records are in the same maximal matching set if and only if they match the same latent individual, though which individual is irrelevant.
Given a set
of records~$\mathcal{A}$, the posterior probability that it is an MMS in $\bm\Lambda$ is simply
%, we can now calculate the posterior probability of a maximal
%matching set which is simply
%
\begin{align*}
P(\bm{\Omega}(\mathcal{A}, \bm{\Lambda}) =1)
%\\
&= 
\frac{1}{S_G}\sum_{h=1}^{S_G}
{\bm{\Omega}(\mathcal{A},\bm{\Lambda}^{(h)})}.
%\sum_{j^{\prime}}{\left( \prod_{(i,j) \in \mathcal{A}}{I(\lambda_{i,j}^{(h)}=j^{\prime})}\prod_{(i,j)\not\in\mathcal{A}}{I(\lambda_{i,j}^{(h)}\neq j^{\prime}})\right)}.
\end{align*}

The MMSs allow a more sophisticated method of preserving transitivity when estimating a single overall linkage structure.
%To describe this method, we first define some additional terminology.
For any record $(i,j)$, its \emph{most probable MMS}~$\mathcal{M}_{ij}$ is the set containing $(i,j)$ with the highest posterior probability of being an MMS, i.e.,
$
\mathcal{M}_{ij}:=
\argmax_{\mathcal{A}:(i,j)\in\mathcal{A}}
P(\bm{\Omega}(\mathcal{A},\bm{\Lambda})=1).
$
Note, however, that there are still problems with a strategy of linking each record to exactly those other records in its most probable maximal matching set.  Specifically, it is possible for different records' most probable maximal matching sets to contradict each other.  For example, Record~A may be in Record~B's most probable maximal matching set, but Record~B may not be in Record~A's most probable maximal matching set.  To solve this problem, we define a \emph{shared most probable MMS} to be a set that is the most probable MMS for each of its members.  We then estimate the overall linkage structure by linking records if and only if they are in the same shared most probable MMS.
The resulting linkage structure is by construction transitive.  
We illustrate examples of MMSs and shared most probable MMSs in Section~\ref{post_match}.

\textcolor{black}{
Finally, we can say that our shared most probably MMSs are optimal under an optimal decision $G^{*},$ i.e., the one that minimizes the posterior expected loss $G^{*} = \arg \min_{G \in \mathcal{G}} W(G),$ where $W(G) = E[L(\lam, G) \mid \bm{X}].$ This was considered in \cite{liseo_2011} under the conference matrix for two files and for the loss functions squared error, false match rate, and absolute number of errors. For our situation, it immediately follows due to \cite{liseo_2011} (Theorem 4.1) that under squared error loss and absolute number of errors, the optimal Bayesian solution is simply for any record $(i,j)$, its \emph{most probable maximal matching set}~$\mathcal{M}_{ij}$ is the set containing $(i,j)$ with the highest posterior probability of being a maximal matching set, i.e.,
$
\mathcal{M}_{ij}:=
\argmax_{\mathcal{A}:(i,j)\in\mathcal{A}}
P(\bm{\Omega}_{\mathcal{A}, \bm{\Lambda}} =1).
$
}
\textcolor{black}{Next, a \emph{shared most probable maximal matching set} is a set that is the most probable maximal matching set of all records it contains, i.e., a set $\mathcal{A}^\star$ such that $\mathcal{M}_{ij}=\mathcal{A}^\star$ for all $(i,j)\in\mathcal{A}^\star$.  We then estimate the overall linkage structure by linking records if and only if they are in the same shared most probable matching set.  The resulting estimated linkage structure is guaranteed to have the transitivity property since (by construction) each record is an element of at most one shared most probable maximal matching set. By Theorem 4.1 of \cite{liseo_2011}, this is the optimal Bayesian decision rule under squared error loss and absolute number of errors. Then trivially under our original definition of the shared MPMMSs, it is still optimal.}

\subsection{Functions of Linkage Structure }
%\textcolor{red}{and Latent Variables}}
\label{sub:function}

%Functions of the linkage structure $\bm{\Lambda}$ can be used to estimate
%quantities such as the number of observed individuals, or summary statistics
%regarding the true attributes of the observed individuals of the population.
%%population.
%Formally, consider any random variable
%%Due to the flexible nature of the linkage structure $\bm{\Lambda},$ we can take functions of this, to estimate quantities of interest, such as the unknown population size or summary statistics regarding true attributes of the population. Suppose there exists an $f$ such that 
%$F = f(\bm{\Lambda}, \bm{X}),$
%where $f: Dom(\bm{\Lambda}) \times Dom(\bm{X}) \rightarrow \mathcal{B},$ 
%and
%%where
%$\mathcal{B}$ is a measurable space.
%Then
%%Hence
%$F \in \sigma(\bm{\Lambda}, \bm{X})$ and $F | \bm{X} = f(\bm{\Lambda}, \bm{X}) | \bm{X}$ can be found from $\bm{\Lambda} | \bm{X}.$ 
%Posterior samples of the linkage structure $\bm{\Lambda}^{(h)}$ imply
%posterior samples for $F$, $\hat{F}^{(h)} = f(\bm{\Lambda}^{(h)},\bm{X}).$
%%Then $\hat{F}_i = f(\hat{\bm{\Lambda}}_i , \bm{X} ).$
%\S \ref{sub:est-function} illustrates how this principle can be used to
%estimate 
%%population
%attributes of observed individuals of the population.
%%Using this flexible notation, as we illustrate in Section \ref{sub:est-function}, we can easily calculate desirable functions of $\bm{\Lambda}$ that applications demand. 

The output of the Gibbs sampler also allows us to estimate the value of any function of the variables, parameters, and linkage structure by computing the average value of the function over the posterior samples.  For example, estimated summary statistics about the population of latent individuals are straightforward to calculate.  Indeed, the ease with which such estimates can be obtained is yet another benefit of the Bayesian paradigm, and of MCMC in particular.

\section{Assessing Accuracy of Matching and Application to NLTCS}
\label{app}

We test our model using data from the National Long Term Care Survey (NLTCS), a
  longitudinal study of the health of elderly (65+)  individuals (\url{http://www.nltcs.aas.duke.edu/}).
  The NLTCS was conducted approximately every six years, with each
  wave containing roughly 20,000 individuals.  Two aspects of the NLTCS make it
  suitable for our purposes: individuals were tracked from wave to wave with
  unique identifiers, but at each wave, 
  many
  %substantial numbers of
  patients had
  died (or otherwise left the study) and were replaced by newly-eligible
  individuals.  We can test the ability of our model to link records across
  files by seeing how well it is able to track individuals across waves, and
  compare its estimates to the ground truth provided by the unique
  identifiers.

To show how little information our method needs to find links across files,
  we gave it access to only four variables, all known to be noisy: date of birth, sex, state of
  residence, and the regional office at which the subject was interviewed.
%To provide a fair illustration of the ability of our method to link records across multiple files,
 % we gave it access to only four variables from each record: date of
 % birth, sex, state of residence, and number of the regional office at which
  %they were interviewed --- all variables known to contain some errors and
 % distortions.
  We linked individuals across the 1982, 1989, and 1994 survey
  waves.
%  \footnote{The 
%  other three waves
%  used different questionnaires and  are not
%  strictly comparable.} 
  Our model had little information on which to link, and not \emph{all} of its
   assumptions strictly hold
%  We provide our model with little information on which to link and not \emph{all}
%  assumptions of the model strictly hold 
 (e.g., individuals can move between states across waves).
%  (e.g. the same individual can reside in different states at different   waves).
%  Thus, we provide our method with comparatively little
%  information on which to link.  Making this more challenging still is that the
%  assumptions of the model Section \ref{model} do not strictly hold here: the same
%  individual can reside in different states at different waves, and the
%  variables are not quite independent (e.g., women's greater longevity
%  correlates sex and year of birth).  
  %
%  Nonetheless, our method
%  does an outstanding job of record linkage, as 
  %
  We demonstrate our method's validity using error rates, confusion matrices, 
  posterior matching sets and linkage probabilities, and estimation of the unknown 
  number of observed individuals from the population.
  %population size. 
  %

Appendix \ref{sec:sim} provides a simulation study of the NLTCS with varying levels of distortion at the field level.  We conclude from this that SMERE is able to handle low to moderate levels of distortion (Figure \ref{distort_six}).  Furthermore, as distortion increases, so do the false negative rate (FNR) and false positive rate (FPR) (Figure \ref{distort}).
  
\subsection{Error Rates and Confusion Matrix}
\label{sub:basic}
Since we have unique identifiers for the NLTCS, we can see how
accurately our model matches records.  A \emph{true link} is a match between
records which really do refer to the same latent individual; a \emph{false link} is a
match between records which refer to different latent individuals; and a \emph{missing
  link} is a match which is not found by the model.  Table~\ref{truelinks_dup}
gives posterior means for the number of true, false, and missing links.
For the NLTCS, 
%we find that
the FNR is $0.11,$ 
%where as
while the FPR is $0.046,$ when we block by date of birth
year (DOB) and sex.

More refined information about linkage errors comes from a confusion matrix,
which compares records' estimated and actual linkage patterns
% estimated linkage patterns to their actual, ground-truth linkage patterns
(Figure \ref{heatmap} and Table~\ref{confusion} in Appendix
\ref{app:confusion}).
% a {\em confusion matrix}, a $(2^k-1) \times (2^k-1)$ contingency table which
% cross-classifies records according to their estimated and ground-truth
% linkage patterns (Table~\ref{confusion}). Diagonal entries of the confusion
% matrix count correct classifications; off-diagonal entries show how often
% mis-classification patterns occur. Mis-classification rates may be calculated
% from the confusion matrix and visualized as a heat map (Figure
% \ref{heatmap}).
%
Every row in the confusion matrix is diagonally dominated, indicating that
correct classifications are overwhelmingly probable.  The largest off-diagonal
entry, indicating a mis-classification, is $0.07$.  For instance, if a record
is estimated to be in both the 1982 and 1989 waves, it is 90\% probable that
this estimate is correct.  
If the true pattern for 1982 and 1989 is wrong, the estimate is most probably 1982 (0.043) and  1989 (0.033) followed by all years (0.018), and and then the other waves with small estimated probabilities.

%all waves (4.4\%), followed by the 1982 wave alone
%(1.4\%) and waves 1982 and 1994 (0.15\%), and then other patterns with still
%smaller probability.}
%\textcolor{blue}{If the estimate is wrong, the truth is most probably
%that the record is in all waves (4.4\%), followed by the 1982 wave alone
%(1.4\%) and waves 1982 and 1994 (0.15\%), and then other patterns with still
%smaller probability.}

%
% For example, we find 118.4 records estimated to be in waves 82 and 89, but
% truly just in wave 82, with a misclassification error of 0.0014.  The heatmap
% illustrates this idea for all combinations of true and estimated linkage
% patterns. The diagonal terms (yellow) indicate correct classifications, which
% are overwhelmingly probable. The off-diagonal terms (orange to dark red)
% indicate how likely an estimated pattern is to be misclassified, where the
% largest rate is 0.07.  To continue with the example, if a record is estimating
% to be in waves 82 and 89, it is 90\% probable that this estimate is correct.
% If the estimate is wrong, however, the truth is most probably that the record
% is in all waves (4.4\%), followed by waves 82 alone (1.4\%) and waves 82 and 94
% (0.15\%), and then other patterns with still less probability.
%(See Table~\ref{confusion:errors} in Appendix~\ref{app:confusion} for detailed calculations.)
\begin{figure}[htbp]
\begin{center}
\includegraphics[width=0.75\columnwidth]{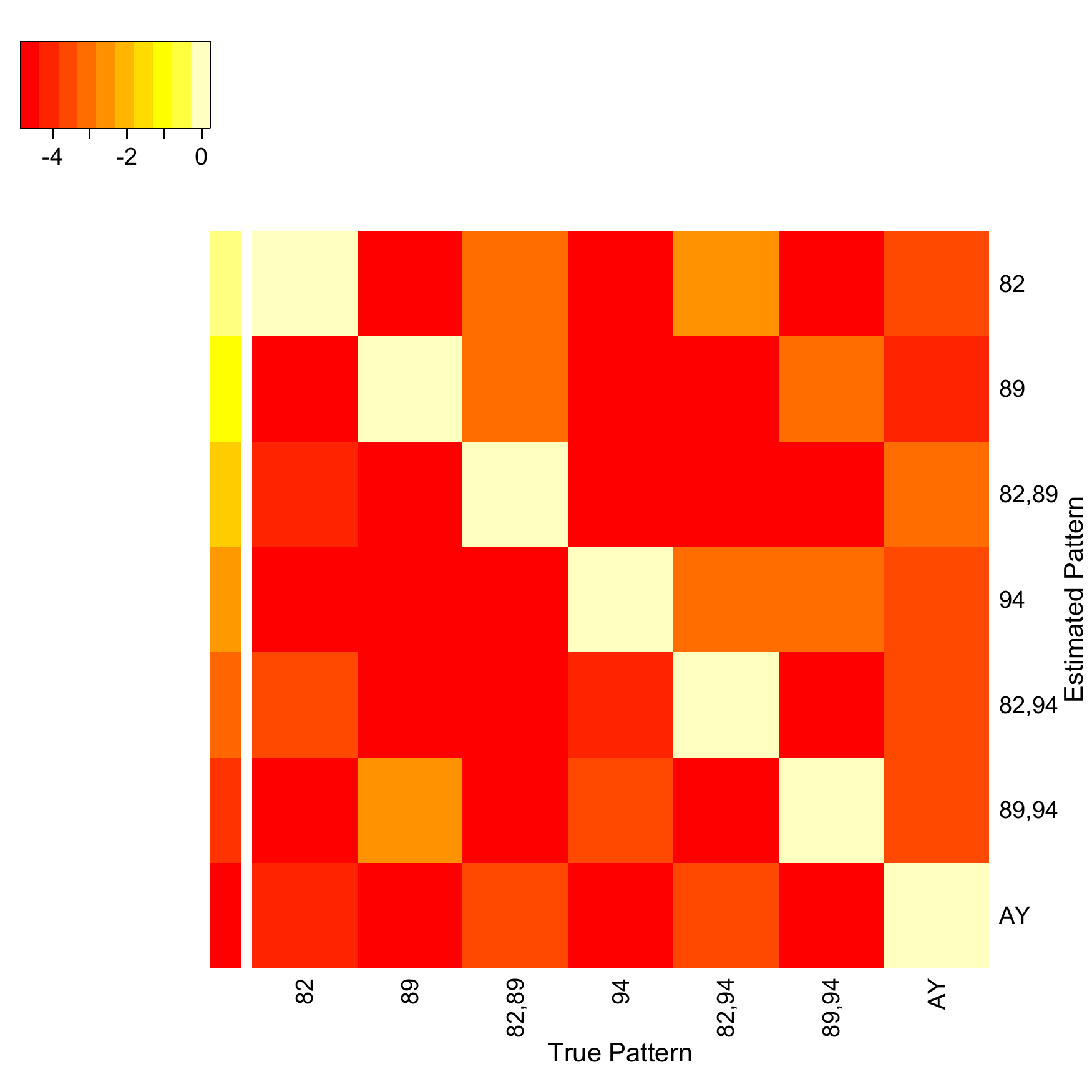}
%\caption{Heatmap of relative probabilities from the confusion matrix, running
%  from yellow (most probable) to dark red (probability 0).  The largest
%  probabilities are on the diagonal, showing that the linkage patterns
%  estimated for records are correct with high probability.  Mis-classification
%  rates are low and show a tendency to under-link rather than over-link.}
\caption{Heatmap of the natural logarithm of the relative probabilities from the confusion matrix, running
 from yellow (most probable) to dark red (least probable). The largest
 probabilities are on the diagonal, showing that the linkage patterns
 estimated for records are correct with high probability. Mis-classification rates are low and show a tendency to under-link rather than over-link, as indicated by higher probabilities for cells above the diagonal than for cells below the diagonal.}
\label{heatmap}
\end{center}
\end{figure}

\subsection{Example of Posterior Matching Probabilities}
\label{post_match}
We wish to search for sets of records that match record 10084 in 1982. In the
posterior samples of~$\bm{\Lambda}$, this record is part of three maximal
matching sets that occur with nonzero estimated posterior probability, one with high and two with low posterior matching probabilities
(Table \ref{postmatch}).  This record has a posterior probability of $0.995$ of
simultaneously matching both record 6131 in 1989 and record 5583 in 1994.  All
three records denote a male, born 07/01/1910, visiting office 25 and residing
in state 14.  The unique identifiers show that these three records are in fact
the same individual.
%If we threshold 
%matching sets, ignoring ones of low posterior probability,
%we would simply return the set of records in last column of
%Table~\ref{postmatch}.
Practically speaking, only matching sets with reasonably large posterior probability, such as the set in the last column of Table~\ref{postmatch}, are of interest.

\subsection{Example of Most Probable MMSs}
%What we do in the code:
%\textcolor{blue}{
%\begin{enumerate}
%\item Create a list of all the records. To do so:
%\begin{enumerate}
%\item Create a lists of all the MMSs from the data (call this all-MMSs)
%\item Separate out each MMS into its component records (and throw out any duplicates that could occur such as over counting).
%\end{enumerate}
%\item Once we have a list of all the records and MPMMS for each record, get all-MPMMs. 
%\item Now we can link records if they share an MPMMS. This can be compared to a baseline.
%\item We can also threshold these MPMMSs (we only link records that share an above thresholded MPMMS). 
%\end{enumerate} 
%}
%\textcolor{blue}{
%So, ignoring the thresholds, being in the same MPMMS is transitive. So if we threshold whole MPMMSs, then transitivity should be preserved. (It is the probabilility that these records will be in the MMS together. This probability must be the same for all records in the same MPMMS. Hence, thresholding, just reduces the number of MPMMSs. It cannot break apart a MPMMS). }

For each record in the NLTCS, we wish to produce its most probable maximal matching set (MPMMS).  We then wish to identify those MPMMSs that are shared.
Finally, we wish to show the linked records for the shared most probable MMSs visually for a subset of the NLTCS (it is too large to show for the entire dataset). 

On each Gibbs iteration, we record all the records linked to a given latent individual; this is an MMS. We aggregate MMSs across Gibbs iterations, and their relative frequencies are approximate posterior probabilities for each MMS. Each record is labeled with the most probable MMS to which it belongs. Finally, we link two records when they share the same most probable MMS, giving us the shared most probable MMS. From this we are able to compute a FNR and a FPR (which we discuss in Section \ref{app:nltcs_dedup}). We give an example of the most probable MMSs for the first ten rows in Table \ref{maximal}.

\begin{table}[h]
\begin{center}
\begin{tabular}{c|cc}
record & MPMMS & posterior probability  \\ \hline
$1.0$ & $\{ 1.0, 1.16651\}$ & $0.51$\\
$1.16651$ & $\{ 1.0, 1.16651 \}$ & $0.51$\\
$1.1$ & $\{ 1.1 \}$ & $1.00$\\
$1.10$ & $\{ 1.10 \}$ & $0.75$\\
$1.2024$ & $\{ 1.2024, 1.21667, 1.39026 \}$ & $0.82$\\
$1.3043$ & $\{ 1.3043 \}$ & $0.58$\\
$1.21667$ & $\{ 1.2024, 1.21667, 1.39026 \}$ & $0.82$\\
$1.39026$ & $\{ 1.2024, 1.21667, 1.39026 \}$ & $0.82$\\
$1.2105$ & $\{ 1.2105, 1.21719, 1.39079 \}$ & $0.81$\\
$1.21719$ & $\{ 1.2105, 1.21719, 1.39079 \}$ & $0.81$
\end{tabular}
\end{center}
\caption{First 10 rows of most probable maximal matching sets. We represent the file and record by $A.B,$ where the file comes first and the record follows the period sign. Hence, 1.10, refers to the tenth record in file one. We use this encoding as it's consistent with our coding practice and easy to refer back to the data in the NLTCS.}
\label{maximal}
\end{table}%

\begin{figure}[htbp]
\begin{center}
\includegraphics{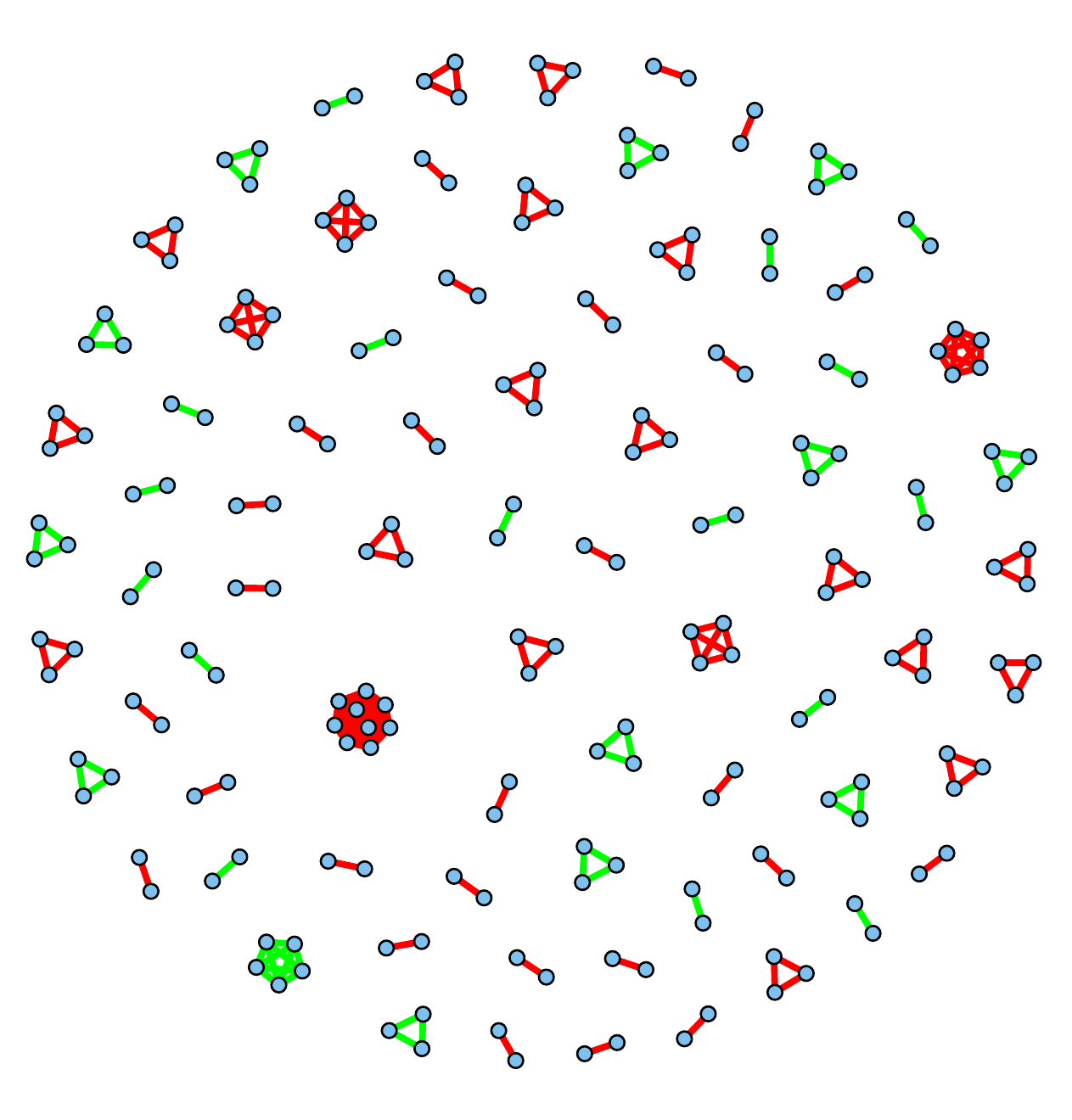}
\caption{This illustrates the first 204 records, where each node is a record and edges are drawn between records with shared most probable MMSs. Transitivity is clear from the fact that each connected component is a clique, or rather, each record is connected to every other record in the same set. Color indicates whether or not the probability of the shared most probable MMS was above (green) or below (red) a threshold of 0.8.}
\label{network}
\end{center}
\end{figure}

\begin{figure}[htbp]
\begin{center}
\includegraphics{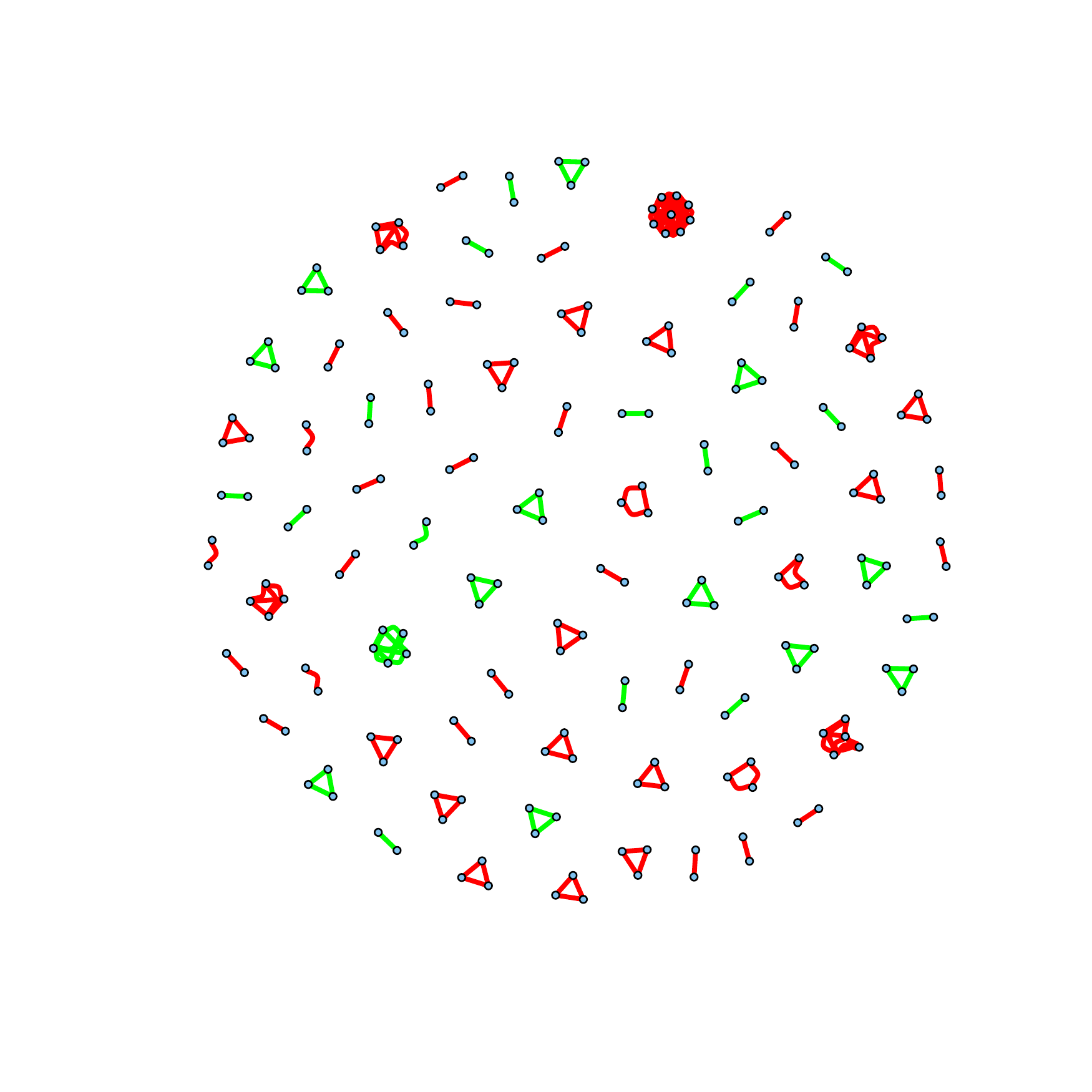}
\caption{The same as Figure \ref{network} with two added features: there are wavy and straight edges. The wavy edges indicate that SMERED and the NLTCS do not agree on the linkage, hence a false link. The straight edges indicate linkage agreement. We can see that there are a fair number of red, wavy edges indicating low probability and incorrect links. There are also many straight edged green high probability correct links. This illustrates one level of accuracy of the algorithm in the sense that it finds correct links and identifies incorrect ones. We can view the red links as individuals we would push to clerical review since the algorithms has trouble matching them (if this is warranted). 
%For the NLTCS, manual review would possibly not do much better than our algorithm since there are many people that match except on unique ID. Since, there is no other information to match on, we cannot hope to improve in this application. 
}
\label{network_curves}
\end{center}
\end{figure}

In Figure \ref{network}, we provide the shared  
most probable MMSs  for the first 204 records of the NLTCS. 
%Color indicates whether or not the probability of a shared most probable MMS was above (green) or below (red), where the thresholded value for coloring applied was 0.8. 
Color indicates whether or not the probability of a shared most probable MMS was above (green) or below (red) the value 0.8.
Transitivity is clear from the fact that each connected component is a clique, i.e., each record is connected to every other record in the same set. 
In Figure \ref{network_curves}, we replot the same records, however, we add a feature to each set. 
Each edge is either straight or wavy to indicate whether the link was correct (straight) or incorrect (wavy) according to the ground truth from the NLTCS. Overall, we see that the the wavy sets tend to be red, meaning that these matches were assigned a lower posterior probability by the model, while the straight sets tend to be green, meaning that these matches were assigned a higher posterior probability by the model.

We can view the red wavy links as individuals we would push to clerical review since the algorithms has trouble matching them. For the NLTCS, manual review would possibly not do much better than our algorithm since there many individuals match on everything except  unique ID. Since there is no other information to match on, we cannot hope to improve in this application.

Useful as these figures are, they would become visually unreadable for the \emph{entire} NLTCS or any record linkage problem of this scale or larger. This is a first step at showing that shared most probable MMSs can be visualized. Visualization on the entire graph structure would be an important advancement moving forward.

%\begin{figure}[htbp]
%\begin{center}
%\includegraphics{pics/network_labels.pdf}
%\caption{The same as Figure \ref{network} except showing the files and the record numbers.}
%\label{network_labels}
%\end{center}
%\end{figure}

\subsection{Estimation of Attributes of Observed Individuals from the Population}
\label{sub:est-function}

\begin{figure}[h]
\begin{center}
\includegraphics[scale = 0.4]{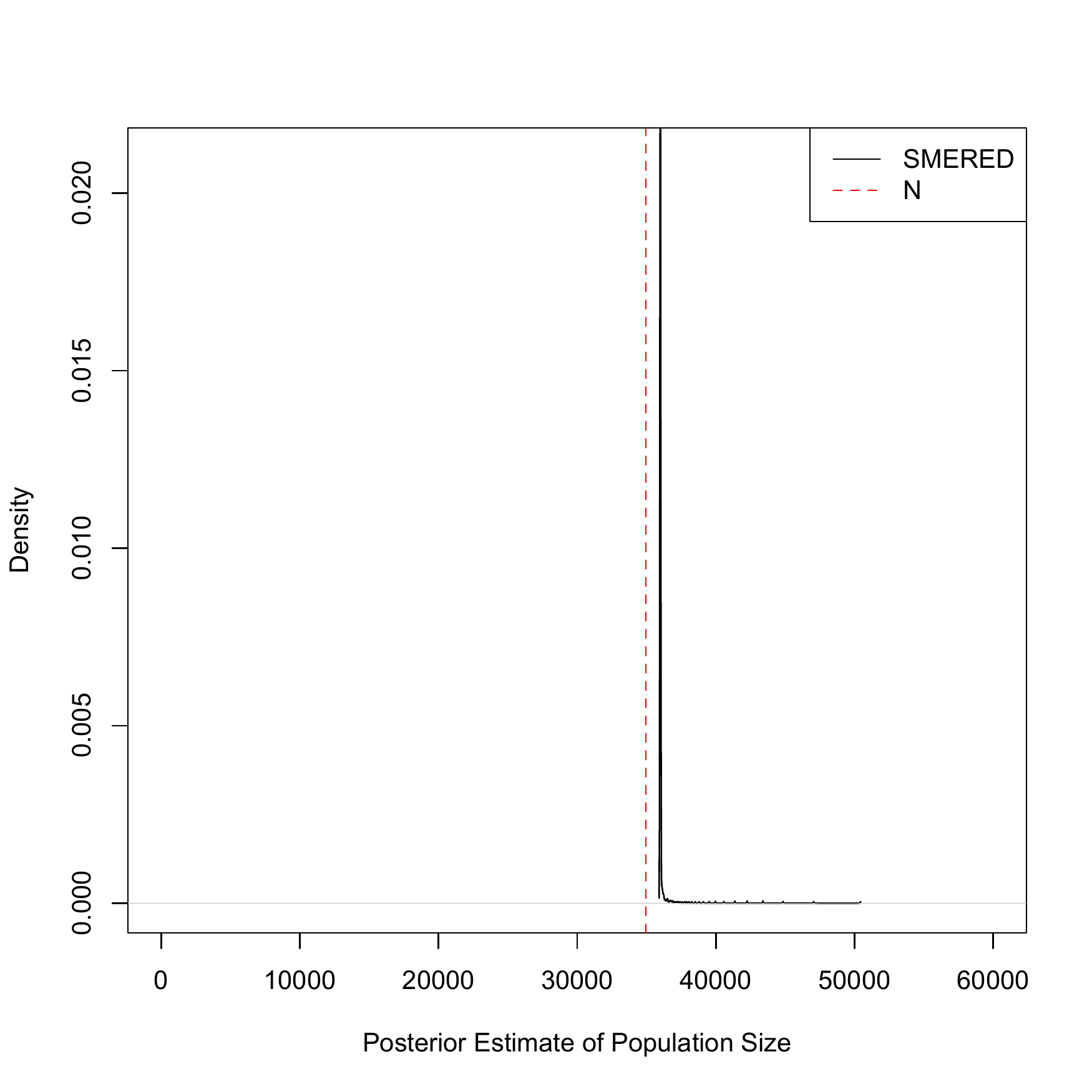}
\caption{Posterior density of the number of observed unique individuals $N$ (black) compared to the posterior mean (red).}
\label{n}
\end{center}
\end{figure}

The number of observed unique individuals $N$ is easily inferred from the posterior of
$\bm{\Lambda}|\bm{X},$ since $N$ is simply the number of unique values in
$\bm{\Lambda}.$
Defining $N|\bm{X}$ to be the posterior distribution of~$N,$ we
can find this by applying a function to the posterior distribution on $\bm{\Lambda}$, 
%this transformation to the posterior distribution
as discussed in Section \ref{sub:function}.
(Specifically, $N=|\#\bm{\Lambda}|$, where $\#\bm{\Lambda}$ maps $\bm{\Lambda}$ to its set of unique entries, and $|A|$ is the cardinality of the set $A$.)
%\textcolor{blue}{Defining $N|\bm{X}$ to be the posterior distribution of~$N,$ we
%can find this by applying this transformation to the posterior distribution
%on~$\bm{\Lambda}.$  (In terms of \S \ref{sub:function}, $f(\bm{\Lambda},\bm{X}) = |\# \bm{\Lambda}|$, where $\# \bm{\Lambda}$ maps $\bm{\Lambda}$ to its set of unique entries, and $|A|$ is the cardinality of the set $A$.)}
Doing so, the posterior distribution of $N|\bm{X}$ is given
in Figure~\ref{n}.  Also, $\hat{N}$:= $E(N|\bm{X}) = 35,992$ with a
posterior standard error of 19.08. The posterior median and mode are 35,993 and 35,982 respectively. 
Since the true number of observed unique individuals is 34,945, we are slightly undermatching, which leads to an overestimate of~$N$. This phenomenon most likely occurs due to
individuals migrating between states across the three different waves. It is
difficult to improve this estimate since we do not have additional information
as described above.  

%\begin{figure}[h]
%\begin{center}
%\includegraphics[scale = 0.4]{pics/posterior_full_run_nostates_heatmaprun2.pdf}
%\caption{Posterior density of the number of observed unique individuals $N.$}
%\label{n}
%\end{center}
%\end{figure}

We can also estimate attributes of sub-groups.
% sub-group instead of sub-populations??
  For example, we can
estimate the number of individuals within each wave or combination of waves---that is, the number of individuals with any given linkage pattern.
%(For brevity,
(We summarize these estimates here with posterior expectations alone, but the
full posterior distributions
are easily computed.)
Recall for each $j'=1,\ldots,N$, $R_{ij'}=\{j:\lambda_{ij}=j'\}.$
For example, the posterior expectation for the number of individuals appearing in
lists $i_i$ and $i_2$ but not $i_3$ is approximately
$$
\frac{1}{S_G} \sum_{h=1}^{S_G} \sum_{j'} 
I\left( \left | R_{i_1j'}^{(h)} \right | = 1 \right)
I\left ( \left | R_{i_2j'}^{(h)} \right | = 1 \right)
I\left( \left | R_{i_3j'}^{(h)} \right | = 0\right ).
$$
(The inner sum is a function of $\bm{\Lambda}^{(h)}$, but a complicated one to express without the $R_{ij}$.)
%, so we omit the full form for brevity and clarity.)

Table~\ref{de-smered_waves} reports the 
posterior means for the
% summary posterior probability for the
overlapping waves and each single wave of the NLTCS and compares this to the
ground truth.
%, which was calculated using a unique person identifier provided in
% the study.
In the first
wave
% year of the study
(1982), our estimates perform
exceedingly well with relative error of 0.11\%, however, as waves cross and we
try to match people based on limited information,
%we see that
the relative errors range from 8\% to 15\%. This is not surprising, since as
patients age, we expect their proxies to respond, making patient data more
prone to errors.  Also, older patients may move across states, creating further
matching dilemmas.
%Without additional information such as name or address, it is difficult to believe that much could be improved upon except modeling different types of distortion or possibly considering
%partial matching instead of exact matching 
We are unaware of any alternative algorithm that does better using this data with
only these fields available.  
%
%Given 
%these results, 
%and considering how little field information we allowed it to use for matching,
%we find that our
%model performs overall very well.

%%new table
%%Comparing NLTCS with 
%%100,000 MH and Gibbs steps (thin = 100, burn-in=1000) and average true-false links. 
%%\vspace*{-1em}
%\begin{table}[htdp]
%\textcolor{red}{Consolidate this table with the later one (now table 4)}
%\begin{center}
%\hspace*{-2em}
%\begin{tabular}{c|ccc|ccc|c}
% & 82& 89 &94& 82, 89 &89, 94& 82, 94  & 82, 89, 94 \\ \hline
%NLTCS (ground truth)  &  7955  & 2959 & 7572
%& 4464  %%82,89
%& 3929  %%89,94
%&  1511  %%82,94
%&  6114 \\ 
%%Bayes$_1$ & 1322 &  8396  & 3430 &8957
%%& 4133 %%82,89
%%&4490 %%89,94
%%& 1625 %%82,94
%%& 5413 \\
%%Bayes$_1$  &  7998.8  & 3473.1 & 8965.6
%%& 4109.4 %%82,89
%%& 4501.6  %%89,94
%%&   1635.9 %%82,94
%%&  5381.8 \\ 
%Bayes  Estimates & 7964.0  & 3434.1 & 8937.8
%& 4116.9%%82,89
%& 4502.1  %%89,94
%&   1632.2 %%82,94
%&  5413.0\\
%Relative Errors (\%)  & 
% 0.11  &16.06 & 18.04 & -7.78 & 14.59 &  8.02 &-11.47
%
%
%\vspace{-.1cm}
%%parametric model &  \\ \hline
%%ground truth & 28,246 & \\
%\end{tabular}
%\end{center}
%%\caption{True and false matches for the parametric model versus the ground truth}
%\caption{Comparing NLTCS (ground truth) to the Bayes estimates of matches }
%\label{linkage-patterns}
%\end{table}

\section{De-duplication}
\label{dedup}

Our application of SMERE to the NLTCS 
assumes
%proceeded under the assumption
that each list
had no
% was free of
 duplicates, however, many other applications will contain duplicates within lists.
We showed in Section \ref{model} that we can theoretically handle de-duplication across and within lists.  We apply SMERE with  de-duplication (SMERED) to the NLTCS by (i) running SMERED on the three waves to show that 
the algorithm does not falsely detect duplicates when there really are none, and (ii) combining all the lists into one file, hence creating many duplicates, to show that SMERED can find them. 
%Furthermore, we create duplicates within each list to access how we handle duplicates in this new setting. We compare these results to our existing results in the previous section. 

%Reinforce message that we can do deduplication across lists and within. We apply this to the NLTCS by combining three files into one lists (such that duplicates are present) and then we deduplicate the files using our SMERED algorithm (computationally more intense). 

\subsection{Application to NLTCS}
\label{app:nltcs_dedup}

We combine the three files of the NLTCS mentioned in Section \ref{app} which contain 22,132 duplicate records out of 57,077 total records.  We run SMERED on settings (i) and (ii),
% as explained
%above,
evaluating accuracy with the unique IDs. We compare our results
 to  ``ground truth" (Table~\ref{de-smered_waves}).  In the case of the NLTCS, compiling all
three files together and running the three waves separately under SMERED
yields similar results, since we match on similar covariate
information. There is no covariate information to add to from thorough
investigation to improve our results, except under simulation
study. 

When running SMERE for three files, the FNR is 0.11 and the FPR is 0.37.  When running SMERED and estimating a single linkage structure by linking records in shared most probable maximal matching sets, the FNR is 0.11 and the FPR is 0.37.  We contrast this with the results obtained when running SMERED under the shared most probable MMSs (MPMMS) for a single compiled file (Table \ref{truelinks_dup}), which yields an FNR of  0.10 and an FPR of 0.17. Clearly, SMERE produces the best results in terms of both FNR and FPR; however, if we want to consider the record linkage and de-duplication problem simultaneously, the SMERED algorithm with linkages applied through the shared MPMMS lowers the FPR nearly in half.

%Specifically, when running SMERED for the three files, the FNR is
%0.11 and is 0.38 for FPR, while its FNR and FPR is 0.11 AND 0.37 for the one
%compiled file. We contrast this with the FNR of 0.11 and FPR of 0.046 under
%SMERE for the three waves (Table \ref{truelinks_dup}).

The dramatic increase in the FPR and number of false links shown in Table
\ref{de-smered_waves} is explained by how few field variables we match on.
Their small number means that there are many records for different individuals
that have identical or near-identical values. On examination, there are $2,558$
possible matches among ``twins,'' records which agree exactly on all attributes
but have different unique IDs.  Moreover, there are 353,536 ``near-twins,'' 
pairs of records that have different unique IDs but match on all but
one attribute.  This illustrates why the matching problem is so hard for the
NLTCS and other data sources like it, where survey-responder information such as
name and address are lacking.  However, if it is known that each file contains
no duplicates, there is no need to consider most of these twins and near-twins
as possible matches.

%\begin{comment}

We would like to put SMERED's error rates in perspective by comparing to another method, but we know of no other that simultaneously links and de-duplicates three or more files.  Thus, we compare to the simple baseline of linking records when, and only when, they agree on all fields \citep[cf.][]{fleming_2007}.
%
%We apply a baseline method for comparison [[need a citation]], since no method of our kind has been proposed for simultaneous linking and de-duplicating three or more files. We compare the baseline to the data run under SMERED.
%
See Table \ref{de-smered_waves} for the relevant error rates. Recall that SMERED produces a FNR and FPR of 0.10 and 0.37. 
%
 %Our baseline is to link records when, and only when, they agree on all fields.  
The baseline has an FPR of $0.09$, much lower than ours, and an FNR of $0.09$, which is the same.  
% Since FNR measures "splitting" of latent individuals, and FPR measures "lumping" of distinct latent individuals, our method is much less prone to lump than the baseline, at a minor cost in splitting. 
We attribute the comparatively good performance of the baseline to there being only five categorical fields per record.  With more information per record, exact matches would be much rarer, and the baseline's FPR would shrink while its FNR would tend to 1.
Furthermore, we extend the baseline to the idea of ``near-twins." Under this, we find that the FPR and FNR are 12.61 and 0.05, where the FPR is orders of magnitude larger than ours, while the FNR is slightly lower. 
While the FPR is high under our model, it is much worse for the baseline of ``near-twins." Our methods tend to ``lump" distinct latent individuals, but it is much less prone to lump than the baseline, at a minor cost in splitting. 
%

%\end{comment}

We also consider a slightly modified version of our proposed methodology wherein we estimate the overall linkage structure by linking those records that are in any shared MMS with a posterior probability of at least 0.8.  The resulting linkage structure is still guaranteed to be transitive, but it includes fewer links.  This procedure attains a similar FNR (0.10) to the original methodology, but it has a substantially lower FPR (0.17).

\begin{figure}[htbp]
\begin{center}
\includegraphics[width=0.8\textwidth]{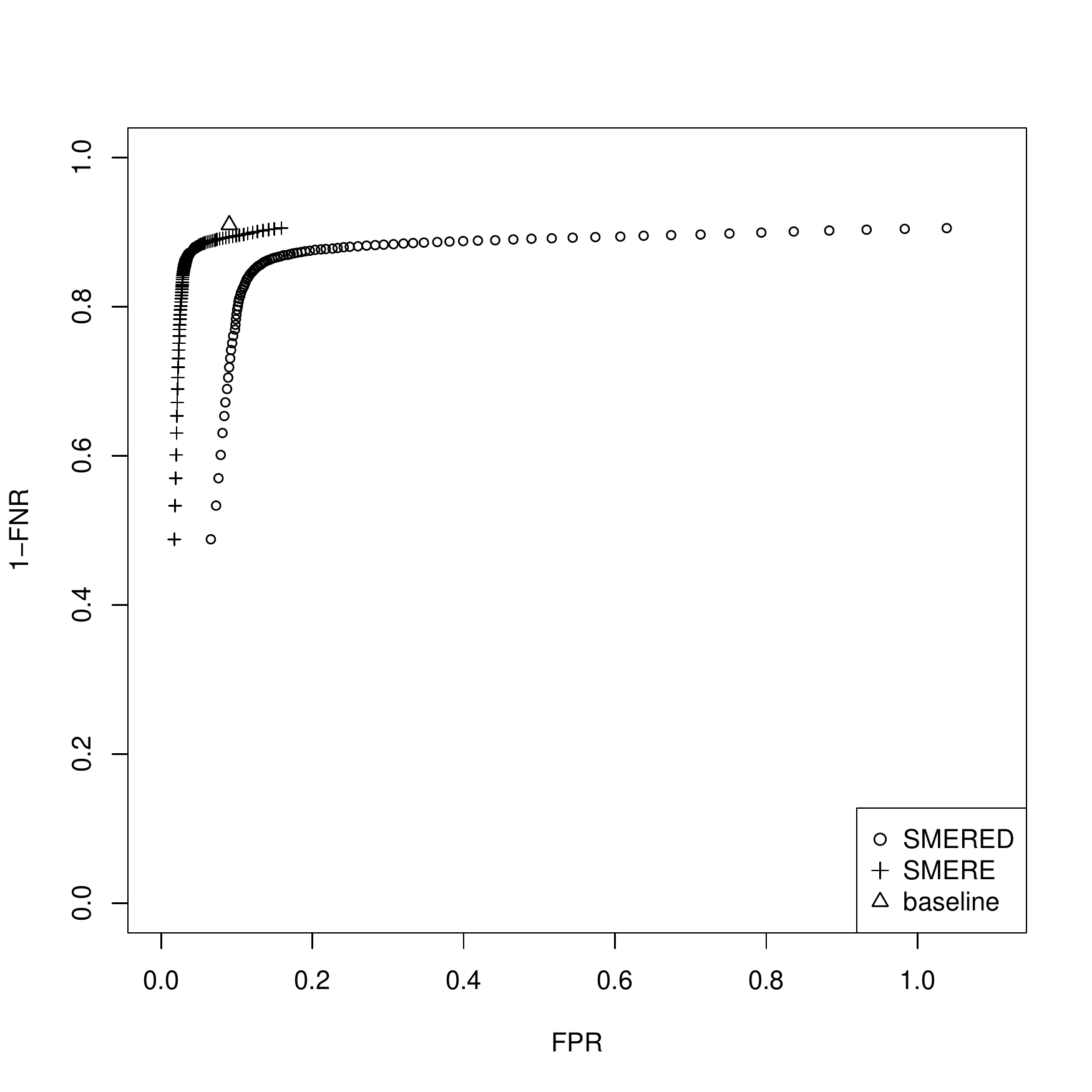}
\caption{We plot both receiver operating characteristics (ROC) curves under SMERE and SMERED for the most probable MMSs (this is to avoid all to all comparisons) and compare them to the simple baseline (triangle). 
For the same FNR ($\approx$ the number of missing links),
the FPR ($\approx$ the number of false links) is higher under SMERED than under SMERE. This is again due to problems in linkage using SMERED when the categorical information is very limited. When the FNR is small, performance is very similar under SMERE and SMERED. However, we note that the baseline can \emph{never change}, whereas, under our algorithm we can relax the FPR or FNR for performance. The ``near-twins" baseline does not appear on this plot since its FPR is 12.61.}
\label{roc_both}
\end{center}
\end{figure}

Furthermore, we compare to the baseline using a range of thresholded FNRs and FPRs using the shared most probable MMSs. We apply thresholded values ranging from [0.2,1] to the shared most probable MMSs (and after thresholding calculating each FNR and FPR). 
This allows us to plot the tradeoff for FNR and FPR under under SMERE and SMERED as seen in Figure \ref{roc_both}. SMERE and SMERED perform similarly compared to the baseline of exact matching, however, we note that either FPR or FNR can be changed for better performance to be gained. When the baseline changes to ``near-twins," however,  both our algorithms in terms of performance radically beat the ``near-twins" baseline, showing the value of a model-based approach 
 when there are distortions or noise in the data. We are not able to catch \emph{all} of the effects of distortion; but, under a very simplistic and easily implemented model, where we match on very little information, we do well.

%\begin{figure}[htbp]
%\begin{center}
%\includegraphics[scale=0.5]{pics/roc_curve_dedupped.pdf}
%\caption{This is the ROC curve under SMERE. }
%\label{roc_smere}
%\end{center}
%\end{figure}
%
%\begin{figure}[htbp]
%\begin{center}
%\includegraphics{pics/roc_curve_dups.pdf}
%\caption{This is the ROC curve under SMERED. }
%\label{roc_smered}
%\end{center}
%\end{figure}

%\begin{figure}[ht]
%\begin{minipage}[t]{0.4\linewidth}
%\centering
%\includegraphics[width=\textwidth]{pics/roc_curve_dedupped.pdf}
%\caption{FNR and FPR plotted against 5 levels of distortion, where the former (triangles) shows near linear relationship and latter shows exponential one (plusses).}
%\label{distort}
%\end{minipage}
%\hspace{0.5cm}
%\begin{minipage}[t]{0.4\linewidth}
%\centering
%\includegraphics[width=\textwidth]{pics/roc_curve_dups.pdf}
%\caption{ROC curve under SMERED and SMERE algorithms.}
%\label{distort_six}
%\end{minipage}
%\end{figure}

\subsection{Application to Italian Household Survey}
\label{app:italy}
We apply our method to the Italian Survey on Household and Wealth (SHIW), a sample survey conducted by the Bank of Italy every two years. The 2010 survey covers 7,951 households composed of 19,836 individuals. The 2008 survey covers  19,907 individuals and 13,266 individuals. We test our methods on all twenty regions, merging the 2008 and 2010 survey. We consider the following categorical variables: year of birth, working status, employment status, branch of activity, town size, geographical area of birth, sex, whether or not Italian national, and highest educational level obtained. We compare our method to that of \cite{liseo_2011} as it is quite natural to our approach. The approach of \cite{liseo_2011} can be framed as a special case of our linkage structure as we describe below. 

Representing a partition as a matrix is not efficient as the number of records $N \rightarrow \infty.$ An alternative is the coreference matrix  \citep{matsakis_10, liseo_2011, sadinle_2014}:
$$\del_{qr} = \begin{cases} 1 &\mbox{if  records $q$ and $r$ refer to same individual}  \\
0 & \mbox{otherwise.}  \end{cases}. $$ We make the relationship clear between that of the coreference matrix of \cite{matsakis_10, liseo_2011, sadinle_2014} and our proposed linkage structure clear.

Lemma: The conference matrix $\del$ can be written as a function of $\lam.$ Hence, the linkage structure can be written or represented as a partition of records. \\

\noindent
Proof: 
Let $q$ and $r$ be two records.
Recall that the coreference matrix is defined as 

$$\del_{qr} = \begin{cases} 1 &\mbox{if  records $q$ and $r$ refer to same individual}  \\
0 & \mbox{otherwise.}  \end{cases} $$
In our notation,  $j$ refers to the individual in list $i.$ Following this notation, for some record $q,$ this corresponds to a list $i_1$ and records $j_1$ in our notation denoted by $(i_1, j_1). $ Similarly, for record $r,$ there is a corresponding list and record denoted by $(i_2, j_2). $
Note: $\lam_{i_1,j_1} = \lam_{i_2,j_2}$ iff $(i_1, j_1)$ and $(i_2, j_2)$ refer to latent individual $j'$ (which has an arbitrary indexing). This implies that 
$\del_{qr} = I(\lam_{i_1,j_1} = \lam_{i_2,j_2}).$ Hence, the coreference matrix can be written as a function of $\lam,$ and thus, $\lam$ can be reordered such that it is a partition matrix. 

Thus, we illustrate that for nearly every region in Italy for the SHIW, SMERED is superior in terms of FNR and FPR to that of \cite{liseo_2011}, something we attribute to the linkage structure, which is easily imbedded within the algorithm of \cite{jain_2004} (See Table \ref{fwiw}).  Our method for every region takes 17 minutes, whereas the competitor approach takes 90 minutes. 

\begin{table}[h]
\begin{center}
\begin{tabular}{ccc|cc}
& Tancredi \& Liseo & &  SMERED \\
Region & FNR  & FPR & FNR & FPR \\ \hline
1 & 1.51 & 0.80 & 0.37 & 0.14\\  
2 & 0.13 & 0.18& 0.34 & 0.02  \\
3 &  0.44 & 0.50 & 0.38 & 0.15\\
4 & 0.45 & 0.47 & 0.62 &
 0.19 \\
5 &  1.26 & 0.75& 0.50 & 0.15\\
6 &  0.54 & 0.58 &0.36 & 0.08\\
7 &  0.70 & 0.58 & 0.42 & 0.10\\
8 &  1.52 & 0.81&  0.43 & 0.17\\
9 & 1.30 & 0.81 & 0.56 & 0.22\\
10 & 0.73 & 0.55 & 0.39 & 0.15\\
11 & 0.97 & 0.63 & 0.54 & 0.18\\
12 & 0.86 & 0.79& 0.40 & 0.12 \\
13 & 0.27 & 0.41 &  0.40 & 0.12 \\
14 & 0.33 & 0.31 & 0.46 & 0.07 \\
15 & 1.11&0.89  & 0.56 & 0.25 \\
16 &0.86 &0.53 & 0.51 & 0.17 \\
17 &0.29 &0.34 &  0.42 & 0.08 \\
18 &0.34 &0.30 & 0.36 & 0.08 \\
19 & 1.18 & 0.71 & 0.50 & 0.16 \\
20 &0.97 &0.54 &  0.45 & 0.11 \\
\end{tabular}
\end{center}
\caption{Method of SMERED versus \cite{liseo_2011}. }
\label{fwiw}
\end{table}%

\section{A User's Guide to Record Linkage}
\label{user}
In this paper, we have proposed a new method and algorithm for simultaneous record linkage and de-duplication. We have focused on two applications, where anonymization occurs, and hence record linkage and de-duplication is useful for such methods. 
%Due to the simplistic nature of our model, we review our assumptions and then give some guidelines for practitioners that might want to use the method. 

We have assumed that (1)  the lists are conditionally independent, (2) the data are categorical, and (3) the records have a minimal set overlap of fields that can be matched.  We then  matched based on our proposed method of MPMMS. We also assumed that blocks are independent and that the distortion probabilities \emph{do not} depend on the list. Moreover, we have used a ``non-infomative" prior on the linkage structure, and we show below that this leads to interesting discoveries and perhaps new research directions in terms of finding subjective priors.
As a user and researcher, one might ask, how valid are these assumptions and what directions are left for exploration?

For applications such as those illustrated here (based on data from the NLTCS and the SHIW), where the data have been  stripped of ``personal identifying information"  for confidentiality reasons, the categorical assumption is in fact valid and very reasonable. For other applications, this assumption is at best an approximation and thus introduces potential errors.  Exploring dependence among lists goes beyond the scope of the present paper and clearly will be context dependent, varying from one problem to the next. The work of \cite{steorts_2015} explores the dependence between lists in an empirically motivated prior setting, showing improvements in the error rates when such dependencies are present. 
%as well as the trade off between adding in a dependence structure and the computational burden created.
%
Moreover,  the applications we have encountered have a common minimal set  of overlapping fields (such as date of birth, location, etc). Situations where some subsets of lists have additional overlap require further study.  These cases and those where there are \emph{no overlapping fields} can be viewed as  missing data problems, e.g., see  \cite{muller_2007}, which proposes a spatial extension of Bayesian Adaptive Regression Trees in a record linkage context.  

Matching to the MPMMS is a good procedure in the sense that it's optimal for squared error loss and the MPMMS preserves transitivity. Thresholding based on this matching criteria is useful in practice since the MPMSS provides a principled way of accounting for the exact uncertainty of transitive matches. It then allows for an automated process to created an updated list of records that match above a posterior probability and push the rest to clerical review. 

Blocking is necessary for dimension reduction and scalability of record linkage models. How to properly block at dataset remains an important issue in record linkage and should be evaluated by the false negative rate. Comparisons of blocking methods have been extensively explored in \cite{steorts_2014_hash}, which includes approaches such as breaking a string into characters, e.g., locality sensitive hashing and random projections. 

The area that needs the most guidance moving forward is in choosing subjective priors
from domain knowledge that are also robust to model-misspecification. We speak to the current limitations of our model and a similar model of \cite{sadinle_2014, sadinle_2015}, which we made connections with earlier.

%\textcolor{red}{Finally, we speak to our choice of a prior on the linkage structure that is ``non-informative," in the sense that it is a Discrete Uniform(1,\ldots, N_{max}). We make two observations here. First, a fixed value of $N_{max}$ is unrealistic and should vary and has recently been explored in \cite{steorts_2014_eb}. Second, we have been able to show that our prior, unsurprisingly along with that of \cite{sadline_2014} yields a highly informative prior on the overall linkage. Recall that in our proposed methods, we link records to a hypothesized latent via the linkage structure. Using this formulation, we can them formulate the prior on the linkage structure instead as a uniform prior on 

\subsubsection*{Assigning Prior Probabilities to Partitions}
Prior probabilities are assigned using $\bm\Lambda$ such that each record is assumed to be equally likely a~priori to correspond to any of the $N$ latent individuals.  Thus, it treats the records as if they are a random sample drawn \emph{with} replacement.  (It is actually perfectly natural to take each of the $N^N$ possible values of $\bm\Lambda$ to be equally likely a~priori).  We now translate this concept to priors on partitions, which requires answering how many possible values of $\bm\Lambda$ correspond to each partition~$\xi$. (Under this translation, we can view both our prior and that of \cite{sadinle_2014} as a prior on partitions).

A partition $\xi$ splits the records up over $|\xi|$ latent individuals, where 
each of these $|\xi|$ latent individuals could have been assigned to one of $N$ index numbers, with the catch that different latent individuals must have different index numbers.
 Thus, there are
$N(N-1)\cdots(N-|\xi|+1)=N!/(N-|\xi|)!$
different $\bm\Lambda$ values that yield the partition $\xi$.
Since each $\bm\Lambda$ value was assigned prior probability $1/N^N$, the prior probability of the partition~$\xi$ is \[
\pi(\xi)=\frac{N!}{(N-|\xi|)!\,N^N}.
\]
Remark: any two partitions with the same number of latent entities are equally probably a~priori.

\subsubsection*{A Generalization of the Probabilistic Structure}

Our prior is constructed by assuming that the $N$ records are randomly sampled with replacement from a population of $N$ latent individuals.  However, why should we assume that the ``population'' of latent individuals has the same size as the sample when considering such a mechanism?  
Since no more than $N$ latent individuals can be represented in the sample, but there 
there \emph{is} a difference between having a latent ``population'' of size $N$ versus something larger than $N$---the larger the latent ``population'' is, the less likely it is for records to be linked.  Instead, let the latent ``population'' have size~$M$.  There are now $M^N$ possible values of $\bm\Lambda$, all of them equally likely.
 There are $M(M-1)\cdots(M-|\xi|+1)=M!/(M-|\xi|)!$ different $\bm\Lambda$ values that yield the partition~$\xi$.
Then the prior probability of the partition~$\xi$ is
\[
\pi(\xi)=\begin{cases}\dfrac{M!}{(M-|\xi|)!\,M^N}&\text{ if }|\xi|\le M,\\0\vphantom{\dfrac00}&\text{ if }|\xi|>M,\end{cases}
\]
noting the impossibility of any partition with more sets ($|\xi|$) then there are individuals in the latent ``population'' ($M$). 

\subsubsection*{The Role of $\bm M$}

The latent ``population'' size $M$ is a tuning parameter that can be altered to control the overall linking tendency.
Suppose $M\to\infty$.  Let $\xi^\star=\{\{1\},\{2\},\ldots,\{N\}\}$ denote the partition in which no records at all are linked, and note that $|\xi^\star|=N$.  Then
\[
\pi(\xi^\star)=\frac{M(M-1)\cdots(M-N+1)}{M^N}=\left(\frac MM\right)\left(\frac{M-1}M\right)\cdots\left(\frac{M-N+1}M\right)\to1
\]
as $M\to\infty$.  If we sample from an essentially infinite population, then our sample is almost guaranteed to consist of $N$ distinct individuals.  However, this is clearly not what we want to do for record linkage.
 On the other end of the spectrum, setting $M<N$ immediately excludes certain linkage structures from consideration.  Specifically, taking $M=N-r$ for some positive integer $r$ implies that all linkage structures with less than $r$ links have prior probability zero.

\subsubsection*{Choosing $\bm M$ via the Prior Mean}

A guideline for choosing $M$ could possibly be determined by looking at the prior expectation of $|\xi|$.  Note that $|\xi|$ represents the number of unique individuals in the sample, so it's a quantity about which we might have some intuition in a particular problem.  Let $\mu(M,N)$ denote this prior mean, i.e., let $\mu(M,N)$ denote the expected cardinality of a random partition $\xi$ of $N$ elements, with the probability of each partition equal to
\[
\pi(\xi)=\begin{cases}\dfrac{M!}{(M-|\xi|)!\,M^N}&\text{ if }|\xi|\le M,\\0\vphantom{\dfrac00}&\text{ if }|\xi|>M.\end{cases}
\]
Note that
\begin{align*}
\mu(M,N)=\sum_{j=0}^{N-1}\left(\frac{M-1}M\right)^j
=M\left[1-\left(\frac{M-1}M\right)^N\right]
%=\frac{M^N-(M-1)^N}{M^{N-1}}
%=\frac{(2M-1)^{N/2}}{M^{N-1}}
&=M\left\{1-\left[\left(1-\frac1M\right)^M\right]^{N/M}\right\}\\
&\approx M\left[1-\left(e^{-1}\right)^{N/M}\right]\\
&=M\left[1-\exp\left(-\frac NM\right)\right]
\end{align*}
as long as $M$ is fairly large. 
% (See the Appendix for a derivation of the first equality above.)
It is not possible to solve  for~$M$ in terms of $\mu(M,N)$.  However, $\mu(M,N)$ is a strictly increasing function of~$M$, so we can pick the value of~$M$ numerically to yield a desired prior mean $\mu(M,N)$. This may be a direction of future research in terms of looking at subjective prior's and partitions. We now illustrate that our prior along with that of \cite{sadinle_2014} is highly informative.

\subsection{Highly Informative Priors}
%
%Everything above \emph{sounds} great.  However, the table above that lists values of $M$ versus $\mu(M,N)$ isn't really telling the whole story.  What's actually happening is that choosing any 
Any
particular value of~$M$ leads to a prior that is still \emph{highly informative} about the overall amount of linkage.  Note that one way to measure the overall amount of linkage is to count the number of individuals represented in the sample.  Both our prior as described above and that of \cite{sadinle_2014} prior are highly informative about this quantity.  For a small sample of size $N=25$, The plot below shows our prior with $M=50$ (blue) and that of \cite{sadinle_2014} prior (red).

\begin{figure}[htbp]
\begin{center}
\includegraphics{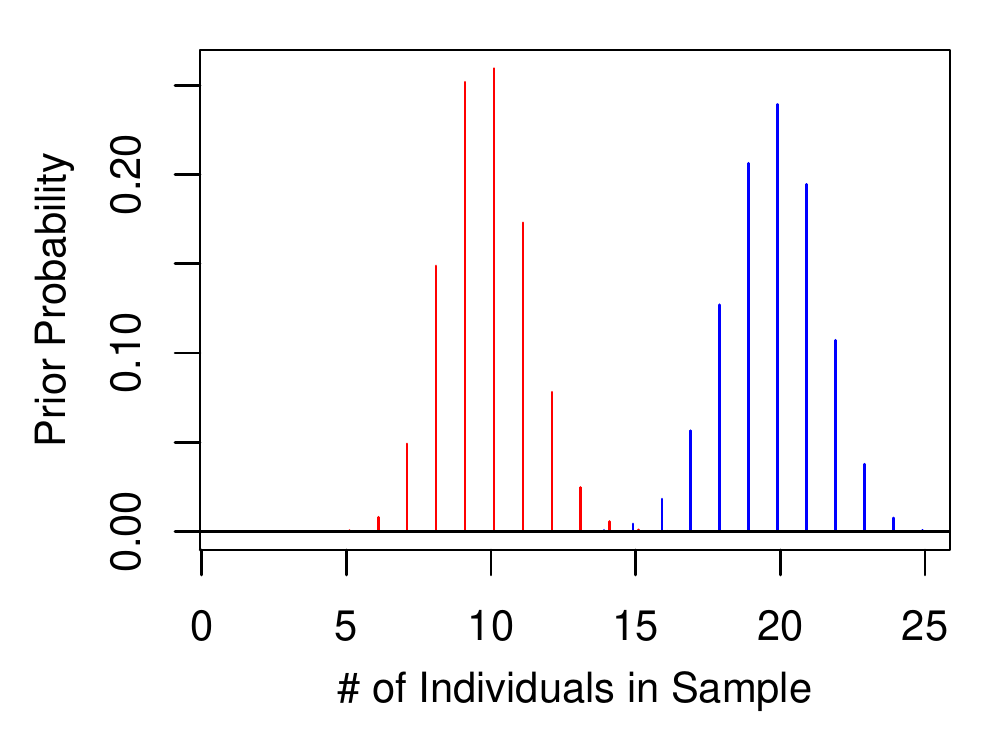}
\caption{Illustrating that when $M=50$ and $N=25$ both the priors of our paper and \cite{sadinle_2014} are highly informative. }
\label{prior}
\end{center}
\end{figure}

Changing $M$ for our prior would shift our distribution to the left or right, but the \emph{location} is not the problem.  The problem is that \emph{the tails of these priors are much too light}.
Moreover, the larger the sample size, the more dramatic this problem will be.
In other words, both priors are highly informative about the overall amount of linkage.  Figure \ref{prior} illustrates that our prior is preferred over that of \cite{sadinle_2014} since the choice of $M$ allows us some control over the situation.  This gives insight at least in the situation of ``non-informative" priors of this type and illustrating how very informative they are! More importantly, this very much drives home the point that we need well principled subjective priors since 
 the light tails we see here mean that the data will not be able to overwhelm the prior if we do not choose $M$ properly (which, in fact, defeats the purpose of being a Bayesian).

%Finally, how do we go about making subjective priors based on domain knowledge at hand? While this topic still needs future work, one suggestion is using simple, fast supervised methods such as random forests to determine possible linkages and use these to elicit part of the prior knowledge as well as better starting points for Bayesian methods that would allow for faster convergence.  

% In summary, the best place to start with evaluation of our record linkage method will work for an application is checking the assumptions stated with the data. Assuming the method is appropriate, then we recommend evaluations such as FNR and FPR and then MPMSS evaluations (as were illustrated in the applications section). 

Hence, it is up to the user to decide whether or not this model and evaluations is appropriate for the data at hand (and to do proper testing as we have suggested) Finally, as we have outlined there is much work left to be done in record linkage and this is a first step at advancing the methodology and understanding the most important steps for moving forward on what is a very important topic both in methods, theory, computer science, and applications. 

\section{Discussion}
\label{disc}
We have made three contributions in this paper.  The first 
and more general one
is to frame record linkage and de-duplication simultaneously, namely
linking observed records to latent individuals and representing the linkage
structure via $\bm{\Lambda}$.  The second contribution is our specific
parametric Bayesian model, which, combined with the linkage structure, allows
for efficient inference and exact error rate calculation (such as our most probable MMS and associated posterior probabilities).  Moreover, this
allows for 
easy
% immediate
integration with capture-recapture methods, where error propagation is
\emph{exact.}
Third, we have suggested practical guidance to practitioners for doing record linkage 
using our proposed methods, outlining its strengths and its shortcomings.
As with any parametric model, its assumptions only apply to certain problems;
%but it also
however, this work
serves as a starting point for more elaborate models, e.g., 
incorporating
missing fields, data fusion, complicated string fields, population
heterogeneity, or dependence across fields, across time, or across individuals.
Within the Bayesian paradigm, 
such model expansions
% these expansions of our model
will lead to
larger
 %higher-dimensional
parameter spaces, and therefore call for computational speed-ups, perhaps via online learning, 
variational inference, or approximate Bayesian computation.

Our work serves as a first basis for
solving
% working with
record linkage and de-duplication problems simultaneously using a noisy Bayesian model and a linkage structure that
can handle large-scale databases. 
We
hope
% expect
that our 
approach
% modeling and our algorithm
will encourage the emergence of new record linkage approaches and applications
% in the social sciences
along with more state-of-the-art algorithms for this kind of high-dimensional
data.

\clearpage
\begin{table}[h]
%\begin{table}[htdp]
\begin{center}
\begin{tabular}{c|ccc}
sets of records & 
 1.10084       &        3.5583; 1.10084  &    3.5583; 1.10084; 2.6131  \\ \hline
posterior probability & 0.001 & 0.004  & 0.995\\
\end{tabular}
\end{center}
\caption{Example of posterior matching probabilities for record 10084 in 1982}
\label{postmatch}
\end{table}

\begin{table}[h]
%\begin{table}[htdp]
\begin{center}
\hspace*{-2em}
\begin{tabular}{c|ccc|ccc|c}
 & 82& 89 &94& 82, 89 &89, 94& 82, 94  & 82, 89, 94 \\ \hline
NLTCS (ground truth)  &  7955  & 2959 & 7572
& 4464  %%82,89
& 3929  %%89,94
&  1511  %%82,94
&  6114 \\ 
%Bayes$_1$ & 1322 &  8396  & 3430 &8957
%& 4133 %%82,89
%&4490 %%89,94
%& 1625 %%82,94
%& 5413 \\
%Bayes$_1$  &  7998.8  & 3473.1 & 8965.6
%& 4109.4 %%82,89
%& 4501.6  %%89,94
%&   1635.9 %%82,94
%&  5381.8 \\ 
Bayes  Estimates$_{\text{SMERE}}$ & 7964.0  & 3434.1 & 8937.8
& 4116.9%%82,89
& 4502.1  %%89,94
&   1632.2 %%82,94
&  5413.0\\
Bayes  Estimates$_{\text{SMERED}}$
 &7394.7  & 3009.9 & 6850.4
& 4247.5%%82,89
& 3902.7  %%89,94
&   1478.7 %%82,94
&  5191.2\\
Relative Errors$_{\text{SMERE}}$  (\%)  & 
 0.11  &16.06 & 18.04 & $-$7.78 & 14.59 &  8.02 & $-$11.47\\
%%TODO: \textcolor{red}{recompute these}
 Relative Errors$_{\text{SMERED}}$  $(\%)$  & 
$-$7.04  & 1.72 &  $-$9.53 & $-$4.85 & $-$0.67 & $-$2.14 & $-$15.09
%0.11 &
%13.83 &
%15.28 &
%$-$8.43 &
%12.73 &
% 7.43 &
% $-$12.95

%\vspace{-.1cm}
%parametric model &  \\ \hline
%ground truth & 28,246 & \\
\end{tabular}
\end{center}
%\caption{True and false matches for the parametric model versus the ground truth}
\caption{Comparing NLTCS (ground truth) to the Bayes estimates of matches for SMERE and SMERED}
\label{de-smered_waves}
\end{table}

\begin{table}[h]
%\begin{table}[htdp]
%\caption{default}
\begin{center}
\begin{tabular}{c|ccc|cc}
& False links & True Links & Missing Links  & FNR & FPR  \\ \hline
NLTCS (ground truth) & 0 & 28246 & 0 & 0 & 0 \\
%Bayes Estimates & 1,282.1 & 25,207.9 & 3,050 \\
Bayes Estimates$_{\text{SMERE}}$ & 1299 & 25196 & 3050 & 0.11 & 0.05\\
%Bayes Estimates (blocking on sex) & 13,564.2 & 13,564.2 & 14,681.8
Bayes Estimates$_{\text{SMERED}}$ &  10595  & 24900  & 3346  &  0.09 & 0.37\\
MPMMSs$_{\text{SMERED}}$ & 4819 & 25489 & 2757 & 0.10 & 0.17\\
Exact matching & 2558 & 25666 & 2580  & 0.09 & 0.09\\
Near-twins matching & 356094 & 26936 & 1310   & 0.05 &12.61\\
%Near-twins matching & 2558 & 25666 & 26976
\end{tabular}
\end{center}
\caption{False, True, and Missing Links for NLTCS under blocking sex and DOB year where the Bayes estimates are calculated in the absence of duplicates per file and when duplicates are present (when combining all three waves). Also, reported FNR and FPR for NLTCS, Bayes estimates.}
% compared to baseline method.}
\label{truelinks_dup}
\end{table}%

%\begin{table}
%%\begin{table}[htdp]
%%\caption{default}
%\begin{center}
%\begin{tabular}{c|ccc}
%&FPR  & FNR  \\ \hline
%NLTCS (ground truth) & 0 & 0 \\
%%Bayes Estimates & 1,282.1 & 25,207.9 & 3,050 \\
%Bayes Estimates$_{\text{SMERE}}$ & 0.11 & 0.05\\
%%Bayes Estimates (blocking on sex) & 13,564.2 & 13,564.2 & 14,681.8
%Bayes Estimates$_{\text{SMERED}}$ &  0.09 & 0.37\\
%Exact matching & 0.09 & 0.09 \\
%Near-twins matching & 0.05 &12.61
%%Near-twins matching & 2558 & 25666 & 26976
%\end{tabular}
%\end{center}
%\caption{False, True, and Missing Links for NLTCS under blocking sex and DOB year where the Bayes estimates are calculated in the absence of duplicates per file and when duplicates are present (when combining all three waves).}
%\label{truelinks_dup}
%\end{table}%

%\textcolor{red}{stick in a table here with the FPR and FNR for each procedure above}

%\twocolumn
\clearpage
\bibliographystyle{ims}
\bibliography{chomp}

\clearpage
%\onecolumn

\appendix

\appendix

\section{Motivating Example of Linkage Structure and Distortion}
\label{example}
We now present a simple example of the ideas of distortion and linkage, 
which illustrates the relationships between the observed data $\bm{X}$, the latent individuals $\bm{y}$, the linkage structure $\lam$, and the distortion indicators $\bm{z}$.
Suppose the ``population'' (individuals
represented in at least one list) has four members, where name and address are stripped for anonymity and they are listed by state, age, and sex.
%
%namely: 
%North Carolina (NC), South Carolina (SC), Pennsylvania (PA), and Virginia (VA).
%Jessica, Margaret, Stephen,and William. 
%
%We represent our latent individuals~($\bm{y}$). Thus,
%$\bm{y}_1$ corresponds to the unobserved values for Jessica, $\bm{y}_2$
%corresponds to the unobserved values for NC and so on. Now suppose our
%lists are written as follows: We know the ground truth here, which is
For instance, the latent individual vector $\bm{y}$ might be 
            \[
         \textcolor{black}{\bm{y}}=
            \left[ {\begin{array}{c}
              \text{NC, 72, F} \\
    \text{SC, 73, F} \\
    \text{PA, 91, M}\\
    \text{VA, 94, M}
                \end{array} } \right]
        .\]

The observed records $\bm{X}$ are given in three separate lists, which would combine into a three-dimensional array.  We write this here as three two-dimensional arrays for notational simplicity:
\[
\text{List 1 }=\begin{bmatrix}\text{NC, 72, F}\\ \text{SC, \textcolor{red}{70}, F}\\\text{PA, 91, M}\end{bmatrix},\qquad
\text{List 2 }=\begin{bmatrix}\text{SC, \textcolor{red}{37} , F}\\
\text{VA, \textcolor{red}{93}, M}\\
\text{PA, \textcolor{red}{92}, M}\end{bmatrix},\qquad
\text{List 3 }=
\begin{bmatrix}\text{NC, 72 , F}\\
\text{NC, 72, F}\\
\text{SC, \textcolor{red}{72}, F}\\
\text{VA, 94, M}
\end{bmatrix}
\]
Here, for the sake of keeping the illustration simple, only age is distorted.

%\begin{figure}[htbp]
% \footnotesize
%  \begin{minipage}{.33\linewidth}
%    \centering
%    \[\text{List 1}=\left[\begin{array}{c}
%    \text{Jessi} \\
%    \text{Margaret} \\
%    \text{William}
%    \end{array}\right]\]
%  \end{minipage}%
%  \begin{minipage}{.33\linewidth}
%    \centering
%    \[\text{List 2}=\left[\begin{array}{c}
%    \text{Marge} \\
%    \text{Steph} \\
%    \text{Bill}
%    \end{array}\right]\]
%  \end{minipage}
%    \begin{minipage}{.33\linewidth}
%    \centering
%    \[\text{List 3}=\left[\begin{array}{c}
%    \text{Jessica} \\
%    \text{Margie} \\
%    \text{Billy}\\
%    \text{Steve}
%    \end{array}\right].\]
%  \end{minipage}
% % \caption{A caption to the entire figure}
%\end{figure}
%%Suppose our list is: List 1 = (Jessi, Margaret, William). \\ List 2~=~(Marge, Steph, Bill). 
%%List 3 = (Jessica, Margie, Billy). \\
%%
%%
%%\vspace*{-1em}
Comparing $\bm{X}$ to $\bm{y}$, the intended linkage and distortions are then
%\[
%\bm{\Lambda}=\begin{bmatrix}1&2&1\\2&3&2\\4&4&4\\&&3\end{bmatrix},\qquad
%\bm{z}=\begin{bmatrix}1&0&0\\1&1&1\\0&1&1\\&&1\end{bmatrix}.
%\]

\vspace*{-2em}
\[
\bm{\Lambda}=\begin{bmatrix} 1 & 2 & 3 &  \\
2 & 4 & 3 &  \\
1 & 1 & 2 & 4 \\
\end{bmatrix},\qquad
\bm{z_1}=  \begin{bmatrix}
0 & 0 & 0\\
0 & 1 & 0 \\
0 & 0 & 0 \\
\end{bmatrix}, \qquad 
\bm{z_2}=  \begin{bmatrix}
0 & 1 & 0\\
0 & 1 & 0 \\
0 & 1 & 0\\\end{bmatrix}, \qquad 
\bm{z_3}=  \begin{bmatrix}
0 & 0 & 0\\
0 & 0 & 0 \\
0 & 1 & 0\\
0 & 0 & 0 \\
\end{bmatrix}.
\]

In this linkage structure, every entry of $\bm{\Lambda}$ with a value of~2 means that some record from~$\bm{X}$ refers to the latent individual with attributes ``SC, 73, F."  Here, the age of this individual is distorted in all three lists, as can be seen from $\bm{z}$.  (Note that $\bm{z}$, like $\bm{X}$, is also really a three-dimensional array.)  Looking at $\bm{z}_1$ and $\bm{z}_3$, we see that there is only a single record in either list that is distorted, and it is only distorted in one field.  In list 2, however, every record is distorted, though only in one field.

\begin{figure}[ht]
\center
%\begin{restofframe}
%\begin{tikzpicture}[>=latex,line join=bevel,]
\begin{tikzpicture}[scale=.45, transform shape,>=latex,line join=bevel,]
\begin{scope}
  \pgfsetstrokecolor{black}
  \definecolor{strokecol}{rgb}{0.0,0.0,0.0};
  \pgfsetstrokecolor{strokecol}
  \draw (8bp,224bp) -- (8bp,442bp) -- (100bp,442bp) -- (100bp,224bp) -- cycle;
  \draw (54bp,424bp) node {$L_1$};
\end{scope}
\begin{scope}
  \pgfsetstrokecolor{black}
  \definecolor{strokecol}{rgb}{0.0,0.0,0.0};
  \pgfsetstrokecolor{strokecol}
  \draw (140bp,80bp) -- (140bp,442bp) -- (234bp,442bp) -- (234bp,80bp) -- cycle;
  \draw (187bp,424bp) node {$L_2$};
\end{scope}
\begin{scope}
  \pgfsetstrokecolor{black}
  \definecolor{strokecol}{rgb}{0.0,0.0,0.0};
  \pgfsetstrokecolor{strokecol}
  \draw (312bp,152bp) -- (312bp,442bp) -- (406bp,442bp) -- (406bp,152bp) -- cycle;
  \draw (359bp,424bp) node {$L_3$};
\end{scope}
\begin{scope}
  \pgfsetstrokecolor{black}
  \definecolor{strokecol}{rgb}{0.0,0.0,0.0};
  \pgfsetstrokecolor{strokecol}
  \draw (135bp,8bp) -- (135bp,60bp) -- (277bp,60bp) -- (277bp,8bp) -- cycle;
\end{scope}
  \node (X_11) at (54bp,394bp) [draw,fill=blue,ellipse] {$X_{11}$};
  \node (X_12) at (54bp,322bp) [draw,fill=blue,ellipse] {$X_{12}$};
  \node (X_13) at (54bp,250bp) [draw,fill=blue,ellipse] {$X_{13}$};
  \node (X_34) at (359bp,178bp) [draw,fill=red,ellipse] {$X_{34}$};
  \node (Y_1) at (170bp,34bp) [draw,fill=purple,ellipse] {$Y_1$};
  \node (Y_2) at (242bp,34bp) [draw,fill=brown,ellipse] {$Y_2$};
  \node (X_21) at (188bp,394bp) [draw,fill=green,ellipse] {$X_{21}$};
  \node (X_23) at (188bp,250bp) [draw,fill=green,ellipse] {$X_{23}$};
  \node (X_22) at (188bp,322bp) [draw,fill=green,ellipse] {$X_{22}$};
  \node (X_25) at (188bp,106bp) [draw,fill=green,ellipse] {$X_{25}$};
  \node (X_24) at (187bp,178bp) [draw,fill=green,ellipse] {$X_{24}$};
  \node (X_31) at (359bp,394bp) [draw,fill=red,ellipse] {$X_{31}$};
  \node (X_32) at (359bp,322bp) [draw,fill=red,ellipse] {$X_{32}$};
  \node (X_33) at (359bp,250bp) [draw,fill=red,ellipse] {$X_{33}$};
  \draw [brown,] (X_21) ..controls (239.44bp,355.23bp) and (292bp,306.13bp)  .. (292bp,250bp) .. controls (292bp,250bp) and (292bp,250bp)  .. (292bp,178bp) .. controls (292bp,129.84bp) and (266.73bp,77.373bp)  .. (Y_2);
  \draw [purple,] (X_22) ..controls (217.67bp,296.4bp) and (229.42bp,282.9bp)  .. (235bp,268bp) .. controls (264.29bp,189.74bp) and (273.85bp,153.98bp)  .. (235bp,80bp) .. controls (227.72bp,66.138bp) and (218.79bp,69.028bp)  .. (206bp,60bp) .. controls (200.21bp,55.91bp) and (193.93bp,51.402bp)  .. (Y_1);
  \draw [purple,] (X_11) ..controls (82.157bp,367.74bp) and (93.907bp,354.24bp)  .. (101bp,340bp) .. controls (119.23bp,303.41bp) and (120bp,290.88bp)  .. (120bp,250bp) .. controls (120bp,250bp) and (120bp,250bp)  .. (120bp,178bp) .. controls (120bp,133.87bp) and (118.85bp,120.66bp)  .. (136bp,80bp) .. controls (140.71bp,68.837bp) and (148.65bp,57.867bp)  .. (Y_1);
  \draw [brown,] (X_34) ..controls (335.16bp,140.21bp) and (318.85bp,104.83bp)  .. (297bp,80bp) .. controls (285.88bp,67.365bp) and (270.67bp,56.253bp)  .. (Y_2);
  \draw [brown,] (X_13) ..controls (65.469bp,197.07bp) and (87.498bp,120.84bp)  .. (136bp,80bp) .. controls (160.75bp,59.16bp) and (176.89bp,74.129bp)  .. (206bp,60bp) .. controls (212.52bp,56.835bp) and (219.11bp,52.489bp)  .. (Y_2);

\end{tikzpicture}
%\end{restofframe}
\caption{Illustration of records $\bm{X},$ latent random variables $\bm{Y},$ and linkage (by edges)~$\bm{\Lambda}$.}
\label{fig:lambda-illustrated}
\end{figure}
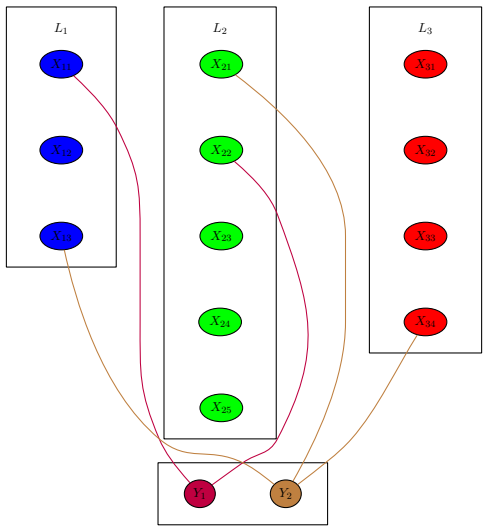

Figure \ref{fig:lambda-illustrated} illustrates the interpretation of our linkage structure as a bipartite graph in which each edge links a record to a latent individual.
For clarity, Figure \ref{fig:lambda-illustrated} shows that $X_{11}$ and
$X_{22}$ are the same individual and shows that $X_{13}, X_{21},$ and $X_{34}$
correspond to the same individual. The rest are non-matches.
%
%The supplementary material (Appendix \ref{sec:linkage-example}) provides a detailed example of linkage structures, distortion, and the meaning of $\bm{\Lambda}$.

\section{Hybrid MCMC Algorithm (SMERED)} \label{app:alg}
\label{algorithm}

We now describe in more detail the Metropolis-within-Gibbs algorithm with
split-merge proposals and optional record linkage blocking (with pseudo-code given at the end).
The entire loop below is repeated for a number of MCMC iterations $S_{H} = S_G \times S_M$.
 Additionally, the algorithm allows multiple split-merge
operations to be performed in a single Metropolis-Hastings
proposal step.  Let $T$ denote the allowed number of split-merge
operations within each Metropolis-Hastings step.  Let
$\bm{\Lambda}^{(m)},\bm{y}^{(m)},\bm{z}^{(m)},\bm{\theta}^{(m)},\bm{\beta}^{(m)}$
denote the values of the MCMC chain at step~$m$.
\begin{enumerate}
\setcounter{enumi}{0}
\item  Repeat the following sequence of steps $T$ times:

\begin{enumerate}
\item 
%Choose a pair of records at random using the \texttt{RandomRecordPair}
%  subroutine. 
  As already described in Section~\ref{alg}, we sample pairs of records from different files uniformly at random within blocks.
    
%
%
%  To choose the pair of records at each stage, one option is to sample
%  uniformly from among all possible pairs.  However there are two problems with
%  this approach.  The first is that we would often select a pair of records
%  which is impossible to merge, since the corresponding individuals are in
%  conflict (e.g., both contain a record in file 1, and hence cannot be merged
%  as that would result in an individual with two records in file 1.  The second
%  reason is that we would often choose pairs of records to merge which were
%  obviously not co-referent (e.g., pairs which agree on no fields whatsoever).
%  To solve the first problem, in our algorithm we simply make sure when we
%  sample a pair for merging that the pair in question can in fact be merged.
%  To solve the second problem we restrict the sampling of record pairs to only
%  those pairs which agree exactly on both sex and birth year, better known as
%  \emph{blocking} in the record linkage literature.  These fields are somewhat
%  reliable, but we give up hope of finding the truly co-referent records which
%  disagree on those fields. This, however, speeds up the algorithm
%  considerably. 

\item If the two records chosen above in~(a) are currently assigned to the same
  latent individual, we propose to split them as follows:  
%  This is done via the \texttt{Split} subroutine, which works as follows:
  \begin{enumerate}
  \item Let $j'$ denote the latent individual to which both records are
    currently assigned.  
  \item Let $C$ denote the set of all \emph{other} records---not including the
    two chosen in step (a) above---who are also assigned to latent
    individual~$j'$.  Note that this set $C$ may be
    empty. 
  \item Give the two records new assignments of latent individuals, calling 
 % using the \texttt{NewIndividualIndex} subroutine.  
 	these new
    assignments $j_1$ and $j_2$.  One of the two individuals stays assigned
    to~$j'$, while the other is assigned to a latent individual
    currently not assigned to any records.

%\textbf{Remark}: Rob, it's not clear to me about precisely what is done here. Can you elaborate?

  \item Randomly assign all the other records in $C$ to either $j_1$ or
    $j_2$, which partitions $C$ into sets $C_1$ and $C_2$.  The inclusion
    of this step is important as the algorithm does not actually actually split
    or merge \emph{records}---it actually splits or merges \emph{latent
      individuals}.  Note that the sets $C_1$ and $C_2$ are designed to
    \emph{include} the two records we chose in step~(a).

  \item The latent individuals $j_1$and $j_2$ get their values
    $\bm{y}_{j_1}'$ and $\bm{y}_{j_2}'$ assigned by simply taking
    them to be \emph{equal} (without distortion) to the exact record values for
    one of the individuals in the sets~$C_1$ and~$C_2$ (respectively), chosen
    at random.
  \item For each record in $C_1$ and $C_2$, the corresponding distortion
    indicators~$z_{ij\ell}'$ for each field are resampled from their respective
    conditional distributions.  Note that some of these may be guaranteed to
    automatically be~1 (whenever a record differs from its corresponding latent
    individual on a particular field).

  \item The above steps generate new proposals~$\bm{\Lambda}'$, $\bm{y}'$, and
    $\bm{z}'$.  We now decide whether to accept or reject the proposal
    according to the Metropolis acceptance probability.  
    %This is carried out by the subroutine \texttt{TestMHAcceptance}. 
     If we accept the
    proposal, then we take $\bm{\Lambda}^{(m+1)}=\bm{\Lambda}'$,
    $\bm{y}^{(m+1)}=\bm{y}'$, and $\bm{z}^{(m+1)}=\bm{z}'$.  If we reject the
    proposal, then we take $\bm{\Lambda}^{(m+1)}=\bm{\Lambda}^{(m)}$,
    $\bm{y}^{(m+1)}=\bm{y}^{(m)}$, and $\bm{z}^{(m+1)}=\bm{z}^{(m)}$.

  \end{enumerate}
\item If instead the two records chosen above in~(a) are currently assigned to
  different latent individuals, we propose to merge them. 
%  using the
%  \texttt{Merge} subroutine.  
  The same basic steps happen: a new state is created in
  which all records which belong to the same individual as either input record
  are all merged into a new individual. The fields for this individual are
  sampled uniformly from these records, distortion variables are all
  re-sampled, and then acceptance probability is tested.

%Rob does not provide a detailed description of \texttt{Merge} but instead simply provides the following description:
%\begin{quote}\slshape
%The same basic steps happen: a new state is created in which all records which belong to the same individual as either input record are all merged into a new individual. The fields for this individual are sampled uniformly from these records, distortion variables are all re-sampled, and then acceptance is tested.
%\end{quote}
%\textbf{Remark} Rob: can you write out a more detailed description of what you mean by the \texttt{Merge} subroutine?
\end{enumerate}
\item Finally, new values $\bm{\theta}^{(m+1)}$ and $\bm{\beta}^{(m+1)}$ are
  drawn from their distributions conditional on the values of
  $\bm{\Lambda}^{(m+1)}$, $\bm{y}^{(m+1)}$, and $\bm{z}^{(m+1)}$ that we just
  selected:
  \[\bm{\theta}^{(m+1)},\beta^{(m+1)}\draw\pi(\bm{\theta},\bm{\beta}\mid\bm{y}^{(m)},\bm{z}^{(m)},\bm{\Lambda}^{(m)},\bm{x}).\]
  % However, as in step~4, there is still some ambiguity about exactly how this
  % is being carried out. Rob, could you please provide a description of your
  % \texttt{ResampleParameters} subroutine?
\end{enumerate}

 \begin{algorithm}[t!]
   \DontPrintSemicolon
     \BlankLine
   \KwData{$\bm{X}$ and hyperparameters}
    Initialize the unknown parameters $\bm{\theta}, \bm{\beta}, \bm{y}, \bm{z},$ and $\bm{\Lambda}.$
   \BlankLine
   \For{$i \leftarrow 1$ \KwTo $S_G$} {
    	\For{$j \leftarrow 1$ \KwTo $S_M$} {\
		\For{$t \leftarrow 1$ \KwTo $S_T$} {\
			Draw records $R_1$ and $R_2$ uniformly and independently at random. \\
			\If{$R_1$ and $R_2$ refer to the same individual}{propose splitting that individual, shifting $\bm{\Lambda}$ to $\bm{\Lambda^{\prime}}$\\
			$r \leftarrow \min{
			\left\{
			 1, \pi(\bm{\Lambda}^{\prime},\bm{y},\bm{z},\bm{\theta},\bm{\beta}|\bm{x}) / \pi(\bm{\Lambda},\bm{y},\bm{z},\bm{\theta},\bm{\beta}|\bm{x}) \right\} 
			 }$}
			\Else{propose merging the individuals $R_1$ and $R_2$ refer to, shifting $\bm{\Lambda}$ to $\bm{\Lambda^{\prime}}$\\
			$r \leftarrow \min{
			\left\{
			 1, \pi(\bm{\Lambda}^{\prime},\bm{y},\bm{z},\bm{\theta},\bm{\beta}|\bm{x}) / \pi(\bm{\Lambda},\bm{y},\bm{z},\bm{\theta},\bm{\beta}|\bm{x}) \right\} } $
			
			}
			Resampling $\bm{\Lambda}$ by accepting proposal with Metropolis probability $r$ or rejecting with probability $1-r.$
			}
	Resample  $\bm{y}$ and $\bm{z}.$
	
	}
	Resample $\bm{\theta}, \bm{\beta}.$
   } 
  \BlankLine
   \Return{$\bm{\theta}|\bm{X}, \bm{\beta}\bm{X}, 
   \bm{y}|\bm{X}, \bm{z}|\bm{X},$ and $\bm{\Lambda}|\bm{X}.$}
   \caption{Split and MErge REcord linkage and Deduplication (SMERED)}
   \label{alg:smered}
 \end{algorithm}
 
\subsection{Time Complexity}
 \label{time}

Scalability is crucial to any record linkage algorithm.
%for large-scale applications.  
Current approaches typically run in polynomial (but super-linear) time in $N_{\max}$. (The method of \cite{sadinle_multi_1} is $O(N_{\max}^k)$, while that of  \cite{domingos_2004} finds the maximum flow in an $N_{\max}$-node graph, which is $O(N_{\max}^3)$, but independent of $k$.)
In contrast, our algorithm is linear in both $N_{\max}$ and MCMC iterations.
%as we now show.

%The total 
Our
running time is proportional to the number of Gibbs iterations $S_G,$ 
so
%hence,
we focus on the time
taken by one
%required for a single
Gibbs step. 
Recall the notation from Section \ref{notation}, and define $M =
\frac{1}{p} \sum_{\ell=1}^p M_{\ell}$ as the average number of possible values
per field ($M \geq 1$).
%Recall that the number of lists is $k$, the number of fields is $p$ and the number of latent individuals is $N \leq N_{\max}$, the total number of records across all lists; let $M =
%\frac{1}{p} \sum_{\ell=1}^p M_{\ell}$ be the average number of possible values
%per field ($M \geq 1$).
The time taken by a Gibbs step
is dominated by
%comes almost entirely from
sampling from the
conditional distributions. Specifically, sampling
$\bm{\beta}$ and $\bm{y}$ are both $O(p N_{\max})$; sampling $\bm{\theta}$ is $O(pMN) + O(pN_{\max}) = O(pMN)$, as is sampling $\bm{z}$.  Sampling $\bm{\Lambda}$ is $O(pN_{\max}M)$ if done carefully. Thus, all these samples can be drawn in 
%the worst case,
time linear in $N_{\max}$. 

%%\theta and beta updated once per Gibbs, while y,z, and lambda updated every MH iteration. 

Since there are $S_M$ Metropolis steps within each Gibbs step and each Metropolis step updates $\bm{y}$, $\bm{z}$, and $\bm{\Lambda}$, the time needed for the Metropolis part of one Gibbs step is $O(S_MpN_{\max}) + O(S_MpMN) + O(S_MpN_{\max}M).$ Since $N \leq N_{\max},$  the run time becomes 
$O(pS_M N_{\max}) + O(MpS_M N_{\max}) = O(MpS_M  N_{\max}). $ On the other hand, the updates for $\bm{\theta}$ and $\bm{\beta}$ occur once each Gibbs step implying the run time is $O(pMN) + O(pN_{\max}).$ Since $N \leq N_{\max},$ the run time becomes $O(pM  N_{\max} + p  N_{\max})  = O(p  M N_{\max} ).$ The overall run time of a Gibbs step is 
$O(pM  N_{\max} S_M) +  O(p  M N_{\max} ) = O(pM  N_{\max} S_M).$ Furthermore, for $S_G$ iterations of the Gibbs sampler, the algorithm is order
$O(pM  N_{\max} S_G S_M).$ If $p$ and $M$ are all much
less than $N_{\max}$, we find that the run time is $O( N_{\max} S_G S_M).$

%We have shown that the algorithm (ignoring blocking), is linear in the number of records and MCMC iterations. 
%Further
%More work is needed to determine how the number of MCMC steps should scale with the number of records and so the total time complexity. 
%We expect this to depend both on how informative the fields are, and on the degree of duplication within and across files.  The effects of blocking on
%run time also deserve careful analysis in future work.

%The scaling of MCMC steps with the number of records 
%%Scalability in terms of the MCMC steps and number of records for computational time complexity 
%deserves careful analysis in future work. 
%We expect this to depend on how informative the fields are, on the degree of duplication within and across files, and blocking. 
%%This may heavily depend on (i) how much field information is available  and (ii) the amount of duplicates present per and across files. The analysis should be refined to take into account blocking procedures. Such considerations are beyond the scope of this paper. 
\textcolor{black}{Another important consideration is the number of MCMC steps needed to produce Gibbs samples that form an adequate approximation of the true posterior.  This issue depends on the convergence properties of the hybrid Markov chain used by the algorithm, which are beyond the scope of the present work.}

\section{Simulation Study}
\label{sec:sim}  
\begin{figure}[h]
\begin{minipage}[t]{0.4\linewidth}
\centering
\includegraphics[width=\textwidth]{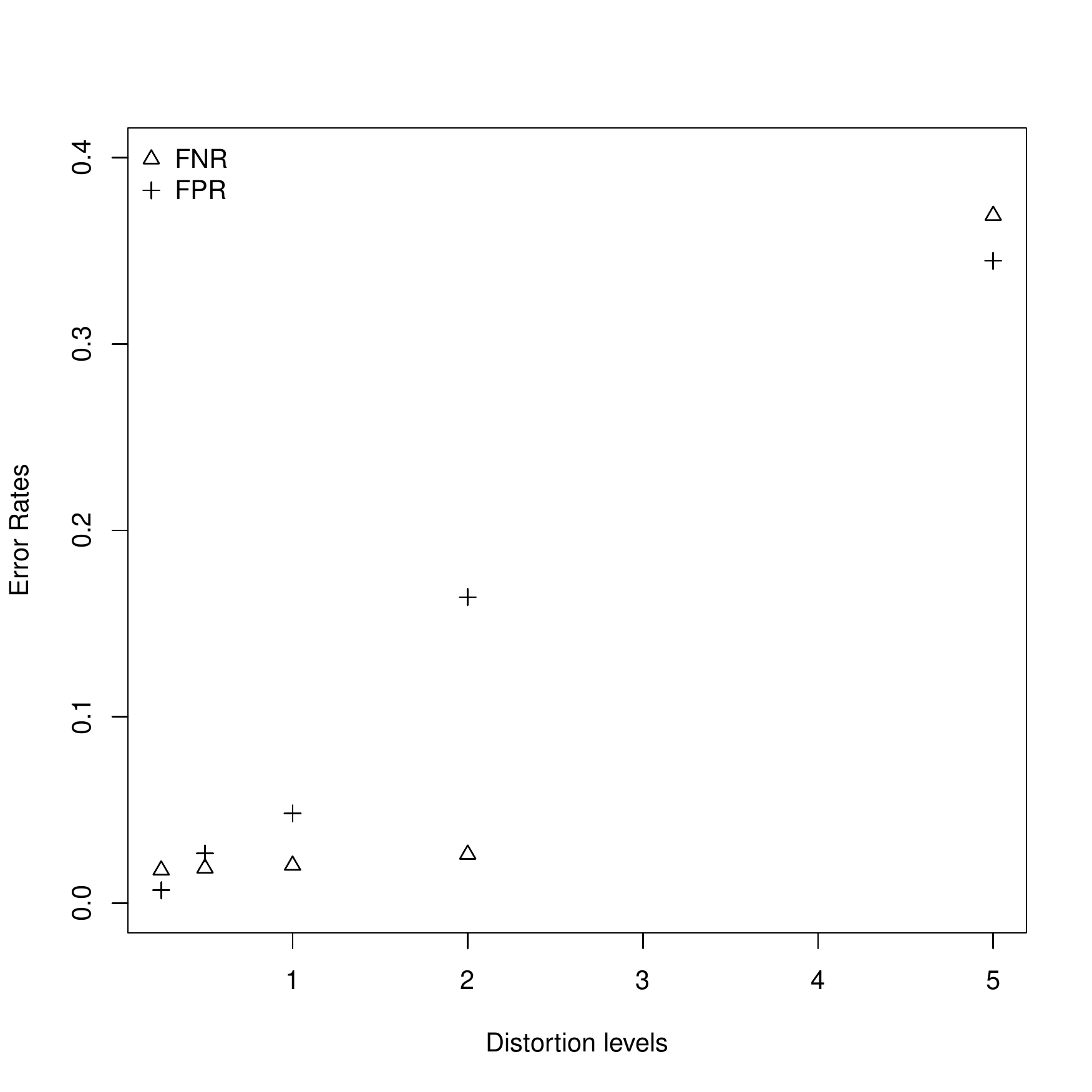}
\caption{
FNR and FPR versus distortion percentage.  FPR shows an approximately linear relationship with distortion percentage, while FNR exhibits a sudden large increase as the distortion level moves from 2\% to 5\%.
%FNR and FPR plotted against 5 levels of distortion, where the former (plusses) shows near linear relationship and latter shows exponential one (triangles).
}
\label{distort}
\end{minipage}
\hspace{0.5cm}
\begin{minipage}[t]{0.4\linewidth}
\centering
\includegraphics[width=\textwidth]{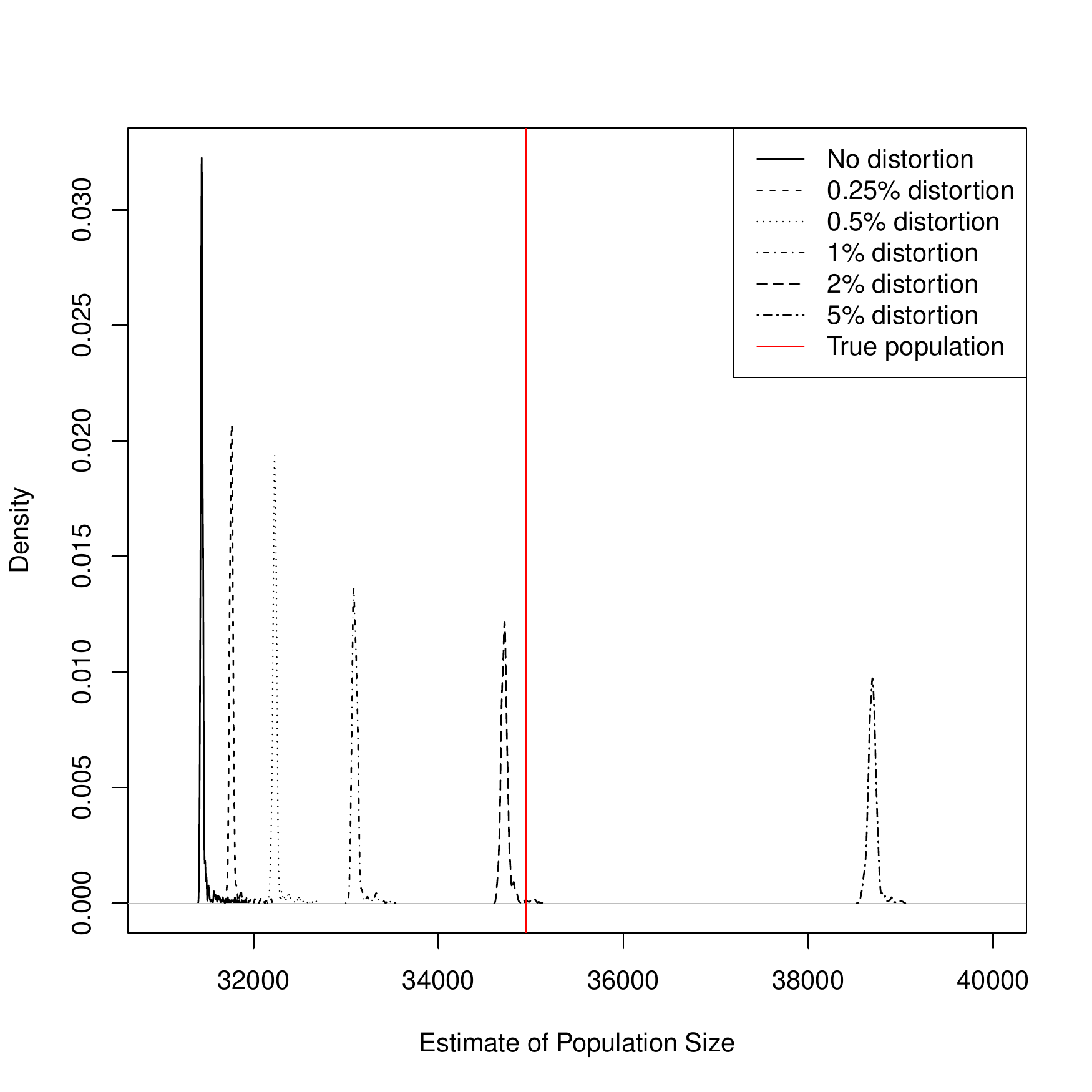}
\caption{Posterior density estimates for 6 levels of distortion (none, 0.25\%, 0.5\%, 1\%, 2\%, and 5\%) compared to ground truth (in red). As distortion increases (and approaches 2\% per field), we overmatch $N$, however as distortion quickly increases to high levels (5\% per field), the model undermatches. This behavior is expected to increase for higher levels of distortion. The simulated data illustrates that under our model, we are able to capture the idea of moderate distortion (per field) extremely well.}
\label{distort_six}
\end{minipage}
\end{figure}

We provide a simulation study based on the model in Section \ref{model}, and we
simulate data from the NLTCS based on our model, with varying levels of
distortion. The varying levels of distortion (0, 0.25\%, 0.5\%, 1\%, 2\%, 5\%)
associated with the simulated data are then run using our MCMC algorithm to
assess how well we can match under ``noisy data.''  Figure \ref{distort}
illustrates an approximate linear relationship with FPR (plusses) and the distortion
level, while for FNR (triangles)
% (triangles),
exhibits a sudden large increase as the distortion level moves from 2\% to 5\%.
%between FNR and the distortion level.
%with FPR and the corresponding distortion level (plusses).
%
Figure~\ref{distort_six} demonstrates that for moderate distortion levels (per
field), we can estimate the true number of observed individuals 
% population size
extremely well via estimated
posterior densities. However, once the distortion is too
\emph{noisy}, our model has trouble recovering  this value.

In summary, 
%we can see that
as records become more noisy or distorted, our matching algorithm
\textcolor{black}{typically matches less than 80\% of the individuals}.  Furthermore, once the
distortion is around 5\%, we can only hope to recover approximately 65\%
of the individuals. 
Nevertheless, this degree of accuracy is in fact quite encouraging given the noise inherent in the data and given the relative lack of identifying variables on which to base the matching.

\clearpage

\section{Confusion Matrix for NLTCS}
\label{app:confusion}

%%this heat map DOES NOT match the confusion matrix
%\begin{table}[htdp]
%\begin{center}
%\hspace*{-2em}
%\small
%\begin{tabular}{c|ccc|ccc|c|c}
%Est vs Truth & 82& 89 &82,89& 94 &82, 94 & 89, 94  & AY & RS\\ \hline
%82 & 8051.9& 0.0& 385.1 &0.0 &162.9 & 0.0& 338.6  & 8938.5\\
%
%89 & 0.0 & 2768.4 & 291.1 & 0.0 & 0.0  & 240.6 & 131.7 & 341.8 \\
%
%94 & 0.0 &  0.0 &  0.0  & 7255.4 & 139.3 & 240.5 & 325.12 & 7960.32 \\
%
%82, 89 & 118.4 & 2.2 & 8071.7 & 0.0 & 4.4  & 0.4 &  803.2 & 9000.3 \\
%
%89, 94&  0.0 & 186.8 & 6.1  & 190.6 & 1.5 & 7365.8 & 488.2 & 8239 \\
%
%82, 94 & 163.1& 0.0 &  9.5& 97.0 & 2662.2 & 0.09 & 331.5 & 3263.39 \\
%
%AY & 62.5 &1.6 &164.4& 28.9 & 51.7 & 10.6 & 15923.7 & 18342.02\\ \hline
%
%%CS & 8395.9 &  2959 &  8927.9/2 &   7571.9 &   3022/2 &  7857.9/2  & 18342/3 \\
% \hline
%NLTCS  &  8396  & 2959 
%&  4464 %%82,89
%& 7572 %%94
%&  1511%%82,94
%& 3929  %%89,94
%&  6114 \\ 
%
%\end{tabular}
%\end{center}
%\caption{Confusion Matrix for NLTCS}
%\label{confusion}
%\end{table}

\begin{table}[h]
\begin{center}
\hspace*{-2em}
\small
\begin{tabular}{c|ccc|ccc|c|c}
Est vs Truth & 82& 89 &82,89& 94 &82, 94 & 89, 94  & AY & RS\\ \hline
82 & 8051.9& 0.0& 385.1 &0.0 &162.9 & 0.0& 338.6  & 8938.5\\

89 & 0.0 & 2768.4 & 291.1 & 0.0 & 0.0  & 240.6 & 131.7 & 341.8 \\

82, 89 & 118.4 & 2.2 & 8071.7 & 0.0 & 4.4  & 0.4 &  803.2 & 9000.3 \\

94 & 0.0 &  0.0 &  0.0  & 7255.4 & 139.3 & 240.5 & 325.12 & 7960.32 \\

82, 94 & 163.1& 0.0 &  9.5& 97.0 & 2662.2 & 0.09 & 331.5 & 3263.39 \\

89, 94&  0.0 & 186.8 & 6.1  & 190.6 & 1.5 & 7365.8 & 488.2 & 8239 \\

AY & 62.5 &1.6 &164.4& 28.9 & 51.7 & 10.6 & 15923.7 & 18342.02\\ \hline

%CS & 8395.9 &  2959 &  8927.9/2 &   7571.9 &   3022/2 &  7857.9/2  & 18342/3 \\
 \hline
NLTCS  &  8396  & 2959 
&  4464 %%82,89
& 7572 %%94
&  1511%%82,94
& 3929  %%89,94
&  6114 \\ 

\end{tabular}
\end{center}
\caption{Confusion Matrix for NLTCS}
\label{confusion}
\end{table}

\begin{table}[h]
\begin{center}
\hspace*{-2em}
\begin{tabular}{c|ccc|ccc|c}
Est vs Truth & 82& 89 &82,89& 94 &82, 94 & 89, 94  & AY \\ \hline
82 & 0.9600 & 0.00000 & 0.04300 & 0.0000 & 0.0540 & 0.0000 & 0.0180 \\ 
89 & 0.0000 & 0.94000 & 0.03300 & 0.0000 & 0.0000 & 3.1e-02 & 0.0072 \\ 
82, 89 & 0.0140 & 0.00074 & 0.90000 & 0.0000 & 0.0015 & 5.1e-05 & 0.0440 \\ 
94 & 0.0000 & 0.00000 & 0.00000 & 0.9600 & 0.0460 & 3.1e-02 & 0.0180 \\ 
82,94 & 0.0190 & 0.00000 & 0.00110 & 0.0130 & 0.8800 & 1.1e-05 & 0.0180 \\ 
89,94  & 0.0000 & 0.06300 & 0.00068 & 0.0250 & 0.0005 & 9.4e-01 & 0.0270 \\ 
AY & 0.0074 & 0.00054 & 0.01800 & 0.0038 & 0.0170 & 1.3e-03 & 0.8700 
\end{tabular}
\end{center}
\label{confusion:errors}
\caption{Misclassification errors of confusion matrix for NLTCS}
\end{table}

%\section{stuff}
%
%
%\input{conditional}

\end{document}